\documentclass{jfm}

\graphicspath{{figures/}}

\usepackage{hyperref}
\usepackage{graphicx}
\usepackage{epstopdf, epsfig}
\usepackage{float}
\usepackage{url}
\usepackage{natbib}
\usepackage{amsmath,amssymb,amstext,mathtools}
\usepackage{scalerel}
\usepackage{stackengine,wasysym}
\usepackage{subcaption}

\def\k{\boldsymbol{k}}
\def\u{\boldsymbol{u}}
\def\dt{\partial t}
\def\D{\mathcal{D}}
\def\P{\mathcal{P}}

\shorttitle{A Nonlocal Spectral Transfer Model for Scalar Turbulence}
\shortauthor{A. Akhavan-Safaei and M. Zayernouri}

\title{A Nonlocal Spectral Transfer Model and New Scaling Law for Scalar Turbulence}

\author{Ali Akhavan-Safaei\aff{1,2}
 \and Mohsen Zayernouri\aff{1,3}\corresp{\email{zayern@msu.edu}}}

\affiliation{\aff{1}Department of Mechanical Engineering, Michigan State University, East Lansing, MI 48824, USA
\aff{2}Department of Computational Mathematics, Science and Engineering, Michigan State University, East Lansing, MI 48824, USA
\aff{3}Department of Statistics and Probability, Michigan State University, East Lansing, MI 48824, USA}

\begin{document}

\maketitle

\begin{abstract}
In this study, we revisit the spectral transfer model for the turbulent intensity in the passive scalar transport (under large-scale anisotropic forcing), and a subsequent modification to the scaling of scalar variance cascade is presented. From the modified spectral transfer model, we obtain a revised scalar transport model using fractional-order Laplacian operator that facilitates the robust inclusion of the nonlocal effects originated from large-scale anisotropy transferred across the multitude of scales in the turbulent cascade. We provide an \textit{a priori} estimate for the nonlocal model based on the scaling analysis of scalar spectrum, and later examine our developed model through direct numerical simulation. We present a detailed analysis on the evolution of the scalar variance, high-order statistics of scalar gradient, and important two-point statistical metrics of the turbulent transport to make a comprehensive comparison between the nonlocal model and its standard version. Finally, we present an analysis that seamlessly reconciles the similarities between the developed model with the fractional-order subgrid-scale scalar flux model for the large-eddy simulation \citep{akhavan2020data} when the filter scale approaches the dissipative scales of turbulent transport. In order to perform this task, we employ a Gaussian process regression model to predict the model coefficient for the fractional-order subgrid model.
\end{abstract}


\section{Introduction} \label{sec: Intro}
~
Understanding the mechanisms responsible for transport of a passive scalar, e.g. temperature field, in high-speed turbulent flow medium is of fundamental importance for scientific and engineering applications. For example, turbulent regime is known to enhance the mixing by the molecular diffusion in passive scalars where it is the result of growth in small-scale fluctuations, distortion of scalar interfaces, and the occurrence of highly intermittent scalar gradients at small scales of transport \citep{warhaft2000passive, shraiman2000scalar, dimotakis2005turbulent, schumacher2005very}. Advancement in understanding of such phenomena is closely dependent on unraveling the complexity that is enforced by the strong nonlinear couplings over a vast range of scales that are also accompanied by the stochastic nature of turbulence \citep{gat1998, frisch1998, sreenivasan2019turbulent}. Therefore, considerable efforts has been devoted to study the structure of turbulent transport in passive scalars in scale-space description at high Reynolds regime \citep{prasad1988multifractal, domaradzki1990local}, especially by focusing on small-scale intermittency and anisotropy effects \citep{kang2001passive, li2006intermittency, watanabe2006intermittency, donzis2008dissipation, donzis2010resolution, kitamura2021spectral}. These efforts were originated from the Kolmogorov's scale-space description of turbulence \citep{kolmogorov1941energy, kolmogorov1941local} that related the statistics of velocity increments to the average dissipation rate of the turbulent kinetic energy (TKE). Kolomogorov's theory was fundamentally constituted based on a local model for turbulent energy cascade as demonstrated in Onsager's cascade model for turbulent spectra \citep{onsager1945distribution, onsager1949}. This theory was later extended to the turbulent transport of the passive scalars by \citep{obukhov1949structure}, \citep{yaglom1949local}, and \citep{corrsin1951spectrum}. Afterwards, the analogy for the different regimes of passive scalar transport given the diffusivity range was developed by \citep{batchelor1959a, batchelor1959b}.

The initial Kolmogorov theory developed in 1941 was later refined in order to take into account the strong intermittency in local energy dissipation rate \citep{kolmogorov1962refinement, oboukhov_1962}. Similarly, highly intermittent fluctuations in the local energy dissipation and also scalar dissipation rates led to the development of the refined similarity hypotheses for passive scalars as presented in \citep{monin_statistical}. 

One of the main pillars of Kolmogorov's theory and its extension to the passive scalars is the local isotropy at small-scale. In the case of passive scalar turbulence with large-scale anisotropy (e.g. non-zero mean gradients), it has been shown that the statistics of small-scale scalar fluctuations remain anisotropic (see e.g., \citep{buaria2021small}) as \citep{pumir1995} showed that for the case of homogeneous shear turbulent flows. On the other hand, passive scalars are known to exhibit anomalies such as the large-scale behavior that cannot be ruled out with the advection-diffusion (AD) equation directly \citep{sreenivasan2019turbulent}. According to \citep{warhaft2000passive}, those occur as a result of anomalous mixing arising from rare events in which a parcel of fluid moves a distance much greater than the integral length scale without equilibrating. In the analogy of Lagrangian path integrals, \citep{shariman_1994} argued that this behavior is identified for a typical fluid path for which the mixing rate is anomalously long rather than for a typical mixing rate but with an atypical path \citep{warhaft2000passive}. This interpretation is an evidence for nonlocal interactions at the large scale levels of turbulent motion originating from the presence of anisotropy. According to \citep{warhaft2000passive}, this behavior is directly linked to the emergence of heavy tails (exponential tails) in the PDF of passive scalar and has been experimentally observed in the turbulent behavior of passive scalars with non-zero mean gradient \citep{gollub_1991, jayesh_1991, jayesh_1992, lane_1993}. A proper approach to account for a mathematical model representing the accumulative source of these nonlocal motions is to revisit the spectral transfer model for the cascade of the passive scalar. In fact, this has been a thriving area of research as reported in different studies such as \citep{hill1978models, sreenivasan1996passive}. A nonlocal spectral transfer model provides a robust link between the large-scale anisotropy at the energy containing range and the universal range throughout the turbulent cascade while accounting for the breakdown of local isotropy at small scales.

According to a comprehensive survey by \citep{suzuki2021fractional}, fractional-order differential operators provide a promising and predictive direction in mathematical modeling of the nonlocal behavior in engineering applications such as mechanics of materials \citep{suzuki2016fractional, yu2016fractional, failla2020advanced, suzuki2021thermodynamically, suzuki2021anomalous, suzuki2021UBT}, modeling the near-wall turbulence \citep{keith2021fractional}, and Reynolds-averaged Navier-Stokes modeling for wall-bounded turbulent flows \citep{mehta2019discovering, song2021variable}, subfilter modeling for large-eddy simulation (LES) of turbulence \citep{samiee2020fractional, di2021two, akhavan2020data, samiee2021tempered}. In particular, LES is known to be an effective technique in computational turbulence research that reduces the computational cost of the simulations by focusing on resolving the larger scales of the transport while the unresolved scales are modeled from the resolved-scale transport quantities \citep{meneveau2000scale, sagaut2006large, moser2020statistical}. From a theoretical point of view, the turbulence closure appearing in the LES equations is the result of applying a general filtering operation to the governing equations. In the convolution kernel $\mathcal{G}_\Delta(\boldsymbol{x}^\prime-\boldsymbol{x})$ for this filtering operation, $\Delta$ is considered to be the arbitrary filter size. In LES, the common practice is to take $\Delta$ large enough towards the intermediate scales of turbulent transport. However, in theory, the filter size could be considered close to the smaller scales of transport ($\eta$) in a way that $\Delta \rightarrow \eta$ \citep{meneveau2000scale}. With this rationale, one can argue that a ``well-resolved'' direct numerical simulation (DNS) that is usually resolved up to $2\eta < \Delta < 5\eta$, may be considered as a candidate for LES with a proper subfilter modeling. Therefore, it is interesting and vital to examine if a developed nonlocal transport model is reconciled with a nonlocal subgrid-scale (SGS) model in terms of \textit{a priori} model identification.

In the present study, we show that using a well-resolved standard DNS data for the transport of a passive scalar with uniform mean-gradient in a moderately high-Reynolds turbulent flow, the 3-D scalar spectrum does not precisely obey the $\k^{-5/3}$ scaling, and follows a scaling that is enforced by the large-scale anisotropy. Utilizing the Corrsin's generalized cascade model \citep{Corrsin1964PoF}, we propose that the modification to the local time-scale associated with the eddies of size $\ell$ in a way to account for the nonlocal interactions. This modification returns a scaling relation that matches the scalar spectrum after parameterization. Subsequently, the total scalar dissipation is revised and an additional term in the form of a fractional Laplacian of the scalar concentration is obtained. The performance of the AD equation that is equipped with this nonlocal term is assessed in a seamless DNS setting. The resulting statistical analysis on the fully-developed turbulent scalar field shows that considering the effects of nonlocal interactions in the mathematical model (AD equation) provides a better a more pronounced prediction of small-scale scalar intermittecny along the direction of large-scale anisotropy, and it provides a consistent scaling for the third-order mixed longitudinal structure function over a wide range of scales. Finally, considering the nonlocal model for the SGS scalar flux proposed in \citep{akhavan2020data}, we study the consistency of that fractional-order SGS model (when $\Delta \approx 2\,\eta$) with the present nonlocal modeling for the spectral transfer in scalar turbulence. Our comparison in terms of parameter identification of models shows a perfect reconciliation between the two modeling approaches with less than 1\% difference. The model identification for the SGS model is obtained from the Gaussian process regression (GPR) trained on the high-fidelity filtered DNS data, while the nonlocal spectral transfer model is calibrated based on the scaling of the 3-D scalar spectrum.

The rest of this work is organized as follows: in section \ref{sec: ScalarTransport}, we introduce the mathematical model for the transport of turbulent passive scalars, their spectral transfer view, and the nonlocal modeling for the spectral transfer. In section \ref{sec: analysis-NL}, we provide a detailed statistical analysis for the nonlocal and standard models in a DNS setting by comparing the single-point, two-point, and high-order small-scale statistical quantities. In section \ref{sec: LES_rec}, the similarities between the current model and the fractional-order subfilter modeling for large-eddy simulation are reconciled. Finally, section \ref{sec: Conclusion} provides the concluding remarks.


\section{Turbulent transport of passive scalars}\label{sec: ScalarTransport}

The Navier-Stokes (NS) equations that govern incompressible fluid flow dynamics are given by
\begin{align}\label{eqn: NS}
    \frac{\partial \u}{\dt} + \u \cdot \nabla \u = -\frac{1}{\rho} \, \nabla p + \nu \, \Delta \u + \boldsymbol{F}; \quad \nabla \cdot \u = 0,
\end{align}
where $\u$ is the velocity field, $\rho$ denotes the density of fluid, $p$ is the pressure, and $\nu$ indicates the kinematic viscosity. Moreover, $\boldsymbol{F}$ represents an external forcing mechanism and in this setting we take it as $\mathcal{A}\,\u$, where $\mathcal{A}$ is a linear indicator function in the spectral domain in order to artificially inject the dissipated TKE into the large scales (low wavenumbers) to generate a statistically stationary isotropic turbulent velocity field (see \citep{akhavan2020parallel}). In order to model the transport of a conserved passive scalar with the diffusivity ($\mathcal{D}$) in this medium, the AD equation is a well-known Fickian mathematical model, which is written in the following form:
\begin{align}\label{eqn: std-AD1}
    \frac{\partial \Phi}{\dt} + \u \cdot \nabla \Phi = \D \, \Delta \Phi.
\end{align}
Reynolds decomposition allows for $\Phi = \langle \Phi \rangle + \phi$, where $\langle \cdot \rangle$ denotes the ensemble-averaging operator, and $\phi$ is the fluctuating part of the scalar field. In the homogeneous turbulence, assumption of a uniform imposed mean gradient, $\nabla \langle \Phi \rangle = \boldsymbol{G}$, is a common practice in order to consider a forcing mechanism for the turbulent intensity (see e.g., \citep{overholt1996direct}). As a result, \eqref{eqn: std-AD1} is rewritten as
\begin{align}\label{eqn: std-AD2}
    \frac{\partial \phi}{\dt} + \u \cdot \nabla \phi = -\boldsymbol{G} \cdot \u + \D \, \Delta \phi.
\end{align}
The governing equations are numerically solved in the standard pseudo-spectral setting for DNS \citep{akhavan2020parallel, overholt1996direct}, and the simulation setup is further explained in section \ref{subsec: discretization}.

Considering $E_\phi(\k,t)$ as the three-dimensional scalar spectrum, spectral budget of the scalar variance reads as
\begin{align}\label{eqn: scalarVar-sp}
    \langle \phi^2 \rangle = \int_0^\infty E_\phi(\k,t) \, d\k,
\end{align}
where $\k$ indicates the wavenumber. Through the assumption of \textit{small-scale isotropy}, the spectral budget for the dissipation rate of scalar variance by molecular diffusion, $\chi$, is given as
\begin{align}\label{eqn: scalarDiss-sp}
    \chi = 2 \, \D \int_0^\infty \k^2 \, E_\phi(\k,t) \, d\k.
\end{align}
Depending on the ratio of $\nu/\D$, and the dissipation rate of TKE ($\varepsilon$), three main wavenumbers are identified for the scalar spectrum:
\begin{align*}
    \k_{\eta_K} \equiv \left(\frac{\varepsilon}{\nu^3}\right)^{1/4}, \quad \k_{\eta_B} \equiv \left(\frac{\varepsilon}{\nu \D^2}\right)^{1/4}, \quad \k_{\eta_{OC}} \equiv \left(\frac{\varepsilon}{\D^3}\right)^{1/4},
\end{align*}
associated with Kolmogorov ($\eta_K$), Batchelor ($\eta_B$), and Obukhov-Corrsin ($\eta_{OC}$) length-scales. Unlike the case with Schmidt number $Sc:=\nu/\D \approx 1$ where all of these three wavenumbers are nearly equal, $\k_{\eta_B}$ and $\k_{\eta_{OC}}$ encode the different behavior of the scalar spectrum in the presence of viscous-convective subrange ($Sc \gg 1$), and inertial-diffusive subrange ($Sc \ll1$), respectively. It is convenient to differentiate the scales of turbulent cascade by the wavenumbers $\k_{EI}$, $\k_{DI}$, as the $\k < \k_{EI}$ represents the \textit{energy-containing} range, and $\k_{EI} < \k$ indicates the \textit{universal equilibrium} range \citep{pope2001turbulent}. Moreover, the universal equilibrium range is split into \textit{inertial-convective} ($\k_{EI} < \k < \k_{DI}$), and \textit{dissipation} ($\k_{DI} < \k $) subranges for the passive scalars with $Sc=1$ \citep{hill1978models}.

\subsection{Scalar spectral transfer and modeling}\label{subsec: spectral-transfer}

Time-evolution for the spectrum of a conserved scalar is governed by (see e.g., \citep{pope2001turbulent, hill1978models})
\begin{align}\label{eqn: std-AD-sp}
    \frac{\partial}{\dt} E_\phi(\k,t) - T(\k,t) = \P(\k,t) - 2 \, \D \, \k^2 \, E_\phi(\k,t),
\end{align}
where $\P(\k,t)$ denotes the spectral content for the large-scale production rate of the scalar variance, and $T(\k,t)$ is the scalar spectral transfer function.
The unknown nature of the $T(\k,t)$ causes a closure problem in \eqref{eqn: std-AD-sp}; thus, a proper modeling for the spectral transfer function is required. $T(\k,t)$ could be defined as the rate of spectral flux function, $F(\k,t)$, per unit wavenumber as (see e.g., \citep{Corrsin1964PoF, hill1978models})
\begin{align}\label{eqn: sp-transfer}
    T(\k,t) \equiv - \frac{\partial F(\k,t)}{\partial \k}.
\end{align}
By integrating \eqref{eqn: sp-transfer} over the 3-D spectral domain, the spectral flux function is obtained.

As a well-known assumption, $\P(\k , t)$ mainly contributes to the energy-containing scales directly, while it is obvious that $\k^2 \, E_\phi(\k,t)$ is mainly considerable in the small scales of the turbulent cascade where diffusion is the dominant transport mechanism \citep{pope2001turbulent}. Therefore, integrating \eqref{eqn: std-AD-sp} over the wavenumber space yields the following:
\begin{align}\label{eqn: sp-balance}
    \frac{\partial}{\dt} \langle \phi^2 \rangle &= \int_0^{\k_{EI}} \left[\P(\k,t)+T(\k,t)\right] \, d\k + \int_{\k_{EI}}^{\k_{DI}} T(\k,t) \, d\k  
    \nonumber \\ &+ \int_{\k_{DI}}^\infty \left[T(\k,t) - 2 \, \D \, \k^2 E_\phi(\k,t)\right] \, d\k 
\end{align}
In the statistically stationary state, the second and third integrals in \eqref{eqn: sp-balance} are approximately zero \citep{pope2001turbulent, hill1978models}. In other words, within the inertial-convective subrange $\partial F(\k)/\partial \k = 0$ by the constant cascade assumption, and for the diffusive subrange $\partial F(\k)/\partial \k = -2 \, \D \, \k^2 \, E_\phi(\k)$. As a result, for the wavenumbers in the inertial-convective subrange $F(\k) = \chi$, integrating with respect to $\k$ and employing \eqref{eqn: scalarDiss-sp}. Subsequently, \eqref{eqn: sp-balance} is rewritten as:
\begin{align}\label{eqn: sp-balance2}
    \frac{\partial}{\dt} \langle \phi^2 \rangle &=
    \underbrace{\P(t)-F(\k_{EI},t)}_{\text{Energy-containing}} + \underbrace{\underbrace{F(\k_{EI},t) - F(\k_{DI},t)}_{\text{Inertial-Convective}} + \underbrace{F(\k_{DI},t)-\chi(t)}_{\text{Dissipation}}}_{\text{Universal Equilibrium Range}}
\end{align}

\begin{figure}
    \begin{center}
        \begin{minipage}[b]{.49\linewidth}
            \centering
            \includegraphics[width=1\textwidth]{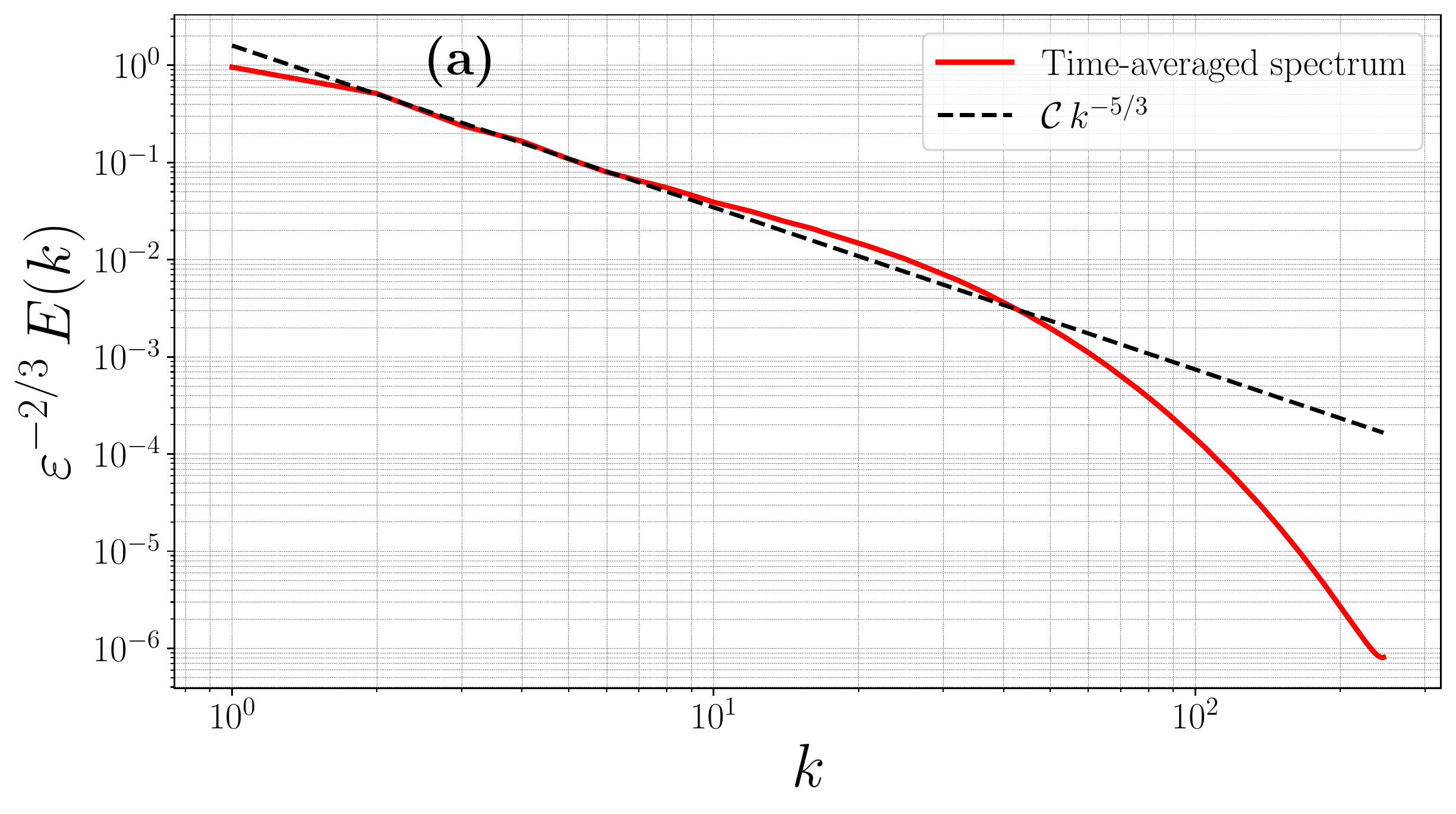}
        \end{minipage}
        \begin{minipage}[b]{.49\linewidth}
            \centering
            \includegraphics[width=1\textwidth]{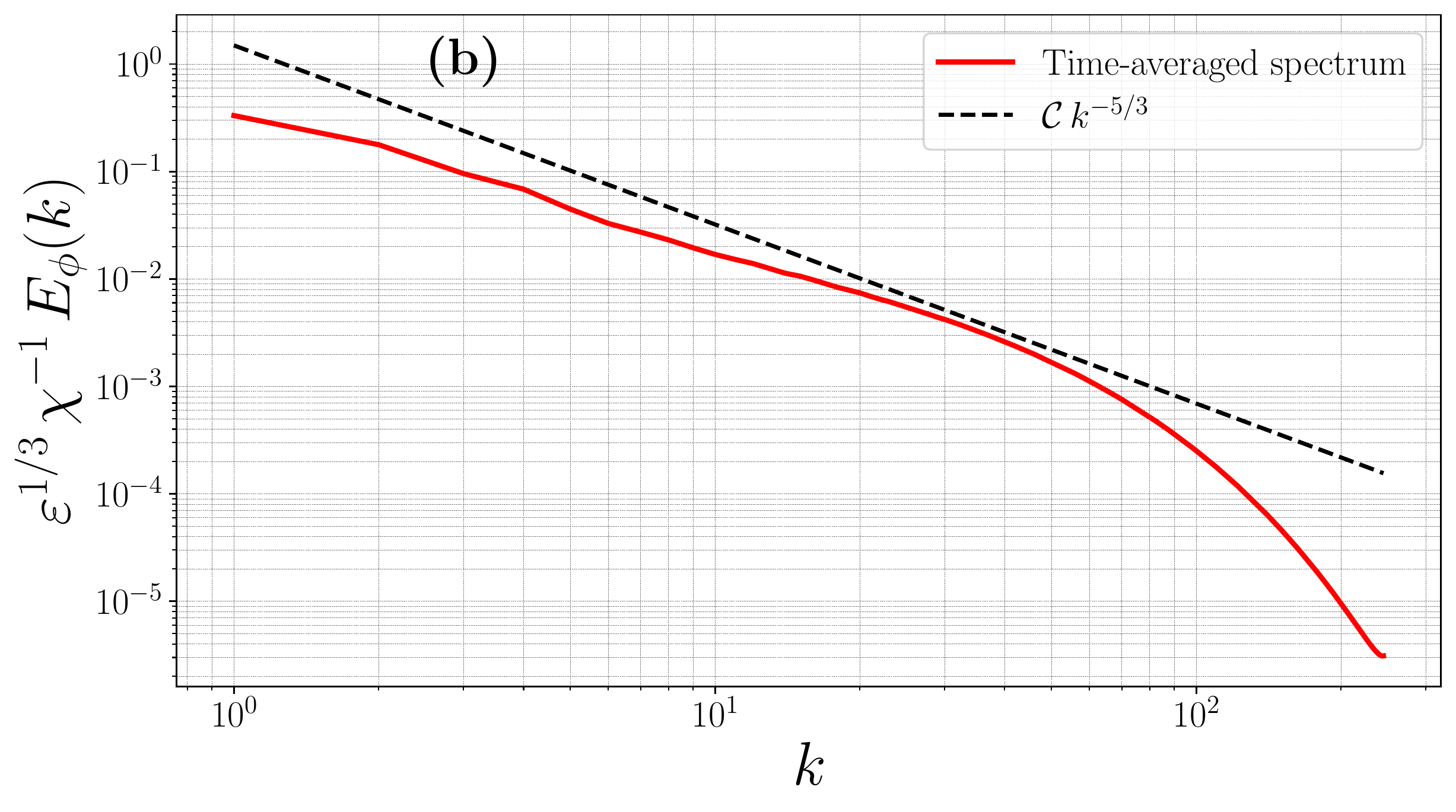}
        \end{minipage}
        \caption{Time-averaged 3-D spectra for (a) turbulent kinetic energy ($E(\k)$), and (b) turbulent scalar intensity ($E_{\phi}(\k)$), obtained from the DNS results described in section \ref{subsec: discretization}.}\label{fig: scaling_TKE-Scalar}
    \end{center}
\end{figure}

This picture motivates the concept of modeling for the turbulent cascade process, here specifically for the case of scalar transport. This approach was initially introduced by Onsager in \citep{onsager1945distribution, onsager1949}, and later was generalized by \citep{Corrsin1964PoF} to the cascade mechanism of other systems such as passive scalars. In the cascade transfer, assuming the \textit{geometric doubling} at each wavenumber step would approximate the step length as $\Delta\k \approx \k$. Therefore, the spectral flux function  could be represented in the following form:
\begin{align}\label{eqn: sp-scalarflux}
    F(\k) \approx \frac{\mathrm{scalar \, variance}}{\mathrm{unit \, time}} = \frac{\Delta\k \, E_\phi(\k)}{\tau(\k)} \approx \frac{\k \, E_\phi(\k)}{\tau(\k)},
\end{align}
where $\tau(\k)$ is the time-scale associated with the step at wavenumber $\k$ \citep{Corrsin1964PoF}. Within the inertial-convective subrange, choosing $\tau(\k)$ to be the local time-scale associated with the eddies of size $\ell=\k^{-1}$, reads as
\begin{align}\label{eqn: local-timescale}
    \tau(\k) = \left( \frac{[\k \, E(\k)]^{1/2}}{1/\k}\right)^{-1} = \left[\k^3 \, E(\k)\right]^{-1/2}.
\end{align}
In \eqref{eqn: local-timescale}, $E(\k)$ is the spectrum of the turbulent kinetic energy. Thus, according to the well-known Kolmogorov's scaling for the inertial subrange, $E(\k) \sim \varepsilon^{2/3} \, \k^{-5/3}$, where $\varepsilon$ is the dissipation rate of TKE.

Plugging \eqref{eqn: local-timescale} into \eqref{eqn: sp-scalarflux}, $F(\k)= \chi \approx \k^{5/2} \, E(\k)^{1/2} \, E_{\phi}(\k)$, and according to the scaling of TKE spectrum, the scalar spectrum scales as
\begin{align}\label{eqn: Sc-sp-scaling1}
    E_\phi(k) &\sim \chi \, \varepsilon^{-1/3} \, \k^{-5/3}.
\end{align}

Direct numerical simulation of the scalar turbulence with a uniform imposed mean gradient, $\boldsymbol{G}=(0,1,0)$, advected on statistically stationary homogeneous isotropic turbulent (HIT) flow, provides a proper database to examine the scaling law in \eqref{eqn: Sc-sp-scaling1}. In this study, a well-resolved DNS at the Taylor-scale Reynolds number $Re_\lambda=240$ for the case with $Sc=1$ is obtained. Resolving over sufficiently large time period in the statistically stationary state provides a rich sample space to obtain the ensemble-averaged spectra for the TKE and scalar. Figure \ref{fig: scaling_TKE-Scalar} illustrates these time-averaged spectra over approximately 15 large-eddy turnover times, $\tau_{ET}$. Although the well-known Kolmogorov scaling for the TKE is achieved as shown in Figure \ref{fig: scaling_TKE-Scalar}a, Figure \ref{fig: scaling_TKE-Scalar}b reveals that the scalar spectrum does not obey the scaling law given in \eqref{eqn: Sc-sp-scaling1}.

\subsection{Nonlocal modeling of the scalar spectral transfer}\label{subsec: spectral-transfer-NL}

\begin{figure}
    \begin{center}
        \includegraphics[width=.7\textwidth]{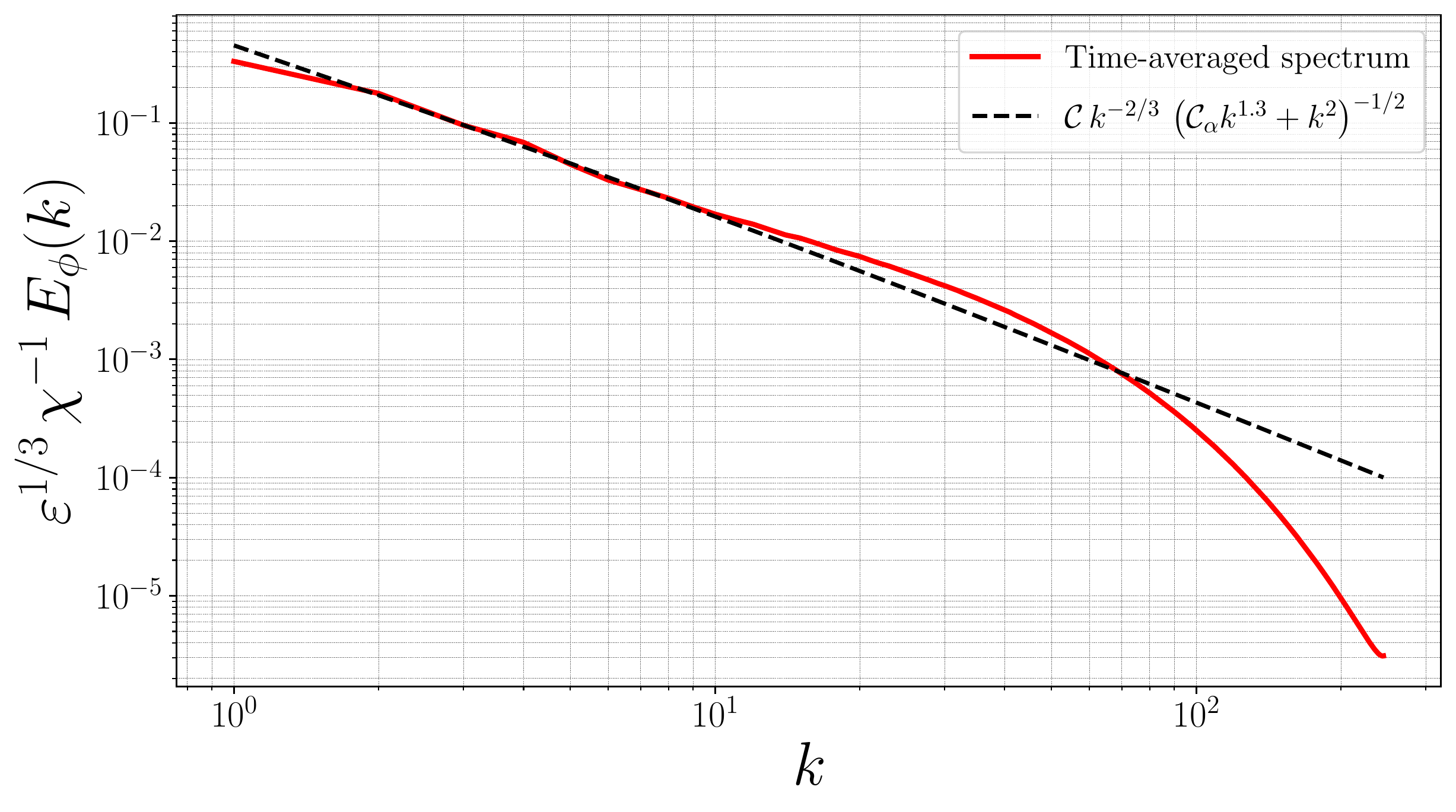}
        \caption{\textit{A priori} identification of the fractional order and $\mathcal{C}_\alpha$, for the modified scaling law introduced in \eqref{eqn: Sc-sp-scaling2} based upon the calibration of the scaling law with the large-scale content of the time-averaged 3-D scalar spectrum ($\k<10$) that induces the nonlocality. The data to compute the scalar spectrum comes from the DNS using standard scalar model. The identified values are $\alpha=0.65$, and $\mathcal{C}_\alpha \approx 3.9$.}\label{fig: nl-Sc-scalaing}
        \end{center}
\end{figure}

In Corrsin's generalization to the Onsager's cascade model, $\partial F/\partial \k$ is considered as the rate of gain or loss of the spectral content per unit wavenumber \citep{Corrsin1964PoF}. Moreover, this generalized model could be applied to

\begin{itemize}
    \item non-conservative systems,
    \item systems with different characteristic time-scales,
    \item systems with different cascading mechanisms.
\end{itemize}

In fact, one can realize that the scaling given in \eqref{eqn: Sc-sp-scaling1} was obtained based upon this generalization. However, in the case of scalar spectrum, a main assumption that might be questioned is small-scale isotropy. Recently, studies showed that effects of large-scale anisotropic forcing do not vanish at the small-scales of turbulent transport of passive scalars. Moreover, these small-scale anisotropic fluctuations are identified to be highly intermittent due to intensified presence of nonlocal interactions in passive scalar turbulence. Anomalous scaling of high-order scalar structure functions is a clear and well-known experimental evidence supporting this significant deviation from local isotropy at small-scales of transport. In fact these effects are available in the inertial-convective subrange. Accordingly, we also observed that the scaling law for the scalar spectrum has a notable discrepancy with the ensemble-averaged spectrum obtained from DNS.

Based on the Corrsin's generalization, we propose that the effects of the anisotropy could be effectively modeled in the spectral transfer function when the nonlocality in the cascade of scalar variance is properly considered. Inclusion of an added power-law behavior into the eddies of size $\ell$ mathematically enables modeling of the long-range interactions (nonlocality), and would return a modification in the local time-scale given in \eqref{eqn: local-timescale}. As a result, we propose
\begin{align}\label{eqn: nl-lengthscale}
    \ell=(\k^2+\mathcal{C}_\alpha \k^{2\alpha})^{-1/2}, \quad \alpha \in (0,1],
\end{align}
where $\mathcal{C}_\alpha$ is a non-negative model coefficient, and $\mathcal{C}_\alpha = 0$ yields the limit case $\ell = \k^{-1}$. Consequently, the modified time-scale is derived as
\begin{align}\label{eqn: nl-timescale}
    \tau(\k) = \left[(\k^2+\mathcal{C}_\alpha \k^{2\alpha}) \, \k \, E(\k)\right]^{-1/2}.
\end{align}
Plugging this nonlocal time-scale into \eqref{eqn: sp-scalarflux} yields the following modified scaling law
\begin{align}\label{eqn: Sc-sp-scaling2}
    E_\phi(\k) &\sim \chi \, \varepsilon^{-1/3} \, \k^{-2/3} \, (\k^2+\mathcal{C}_\alpha \k^{2\alpha})^{-1/2}.
\end{align}
Testing this modified scaling law on the time-averaged scalar spectrum shows a promising agreement through proper parameterization of $\alpha$ and $\mathcal{C_\alpha}$. Figure \ref{fig: nl-Sc-scalaing} illustrates that with $\alpha = 0.65$ and $\mathcal{C}_\alpha\approx 3.9$ the universal scaling observed in the TKE spectrum is achieved for the scalar spectrum with respect to \eqref{eqn: Sc-sp-scaling2}. 

The multi-scale nature of the turbulent cascade process implies that the nonlocal transport effects induced by the large-scale anisotropy are fundamentally connected to the small-scales of motion through inter-scale interactions. Here, the inertial-convective subrange essentially plays the role of a meso-scale region for the turbulent cascade, where the presence of these nonlocal inter-scale interactions are highly pronounced. In fact, several fundamental studies focused on these nonlacal interactions and tried to unravel their nature by triad and tetrad dynamic models in the spectral domain \citep{chertkov1995normal, chertkov1999lagrangian, waleffe1992nature, cheung2014exact}. Therefore, a modification to the dissipation model \eqref{eqn: scalarDiss-sp} (initiated from the small-scale isotropy hypothesis) would compliment the nonlocal effects observed in larger scales of turbulent cascade. Subsequently, the total dissipation ($\chi_T$) is revised into
\begin{align}\label{eqn: scalarDiss-sp-nl}
    \chi_T &= 2 \, \D \int_0^\infty [\k^2 + \mathcal{C}_\alpha \, \k^{2\alpha}] \, E_\phi(\k,t) \, d\k \nonumber \\
         &= \underbrace{2 \, \D \int_0^\infty \k^2 \, E_\phi(\k,t) \, d\k}_{\chi} \, + \, \underbrace{2 \, \D \, \mathcal{C}_\alpha \int_0^\infty \k^{2\alpha} \, E_\phi(\k,t) \, d\k}_{\chi_\alpha}.
\end{align}
Defining $\D_\alpha := \D \, \mathcal{C}_\alpha$, $\chi_\alpha$ characterizes the essence of having an underlying nonlocal diffusion operator in the advection-diffusion equation governing the turbulent transport of passive scalar. From a mathematical point of view, $\chi_\alpha$ directly stems from a fractional-order Laplacian operator, $(-\Delta)^\alpha(\cdot)$, acting on the scalar concentration; thus, the modified transport model reads as  
\begin{align}\label{eqn: std-AD-nl}
    \frac{\partial \phi}{\dt} + \u \cdot \nabla \phi = -\boldsymbol{G} \cdot \u + \D \, \Delta \phi + \D_\alpha \, (-\Delta)^\alpha \phi.
\end{align}

\subsection{Numerical discretization and simulation details}\label{subsec: discretization}

A standard pseudo-spectral scheme based on the Fourier-collocation method is utilized to discretize and resolve the NS and AD equations. The triply periodic computational domain $\boldsymbol{\Omega}=[0,2\pi]^3$ is discretized in space by a uniform grid with $520^3$ grid points. A fourth-order Runge-Kutta (RK4) scheme is employed to perform the time-integration with a fixed $\Delta t=8\times10^{-4}$, while the $\mathrm{CFL} < 1$ condition is always checked; therefore, the numerical stability is ensured. In this study, we select a fully developed HIT field at $Re_\lambda=240$ as the initial state of our computational experiment, and investigate the development of the passive scalar concentration under the effect of a large-scale uniform imposed mean-gradient, $\boldsymbol{G}=(0,1,0)$. Concentration field, $\phi(\boldsymbol{x})$, is initiated from zero and is resolved for approximately 30 large-eddy turnover times under the standard model \eqref{eqn: std-AD2}, and the introduced nonlocal model \eqref{eqn: std-AD-nl}, while $Sc=1$. More detailed discussions about the numerical method, computational approach for the simulations, and the flow characteristics of the utilized HIT data have been presented in \citep{akhavan2020parallel}.

\section{Statistical analysis of the nonlocal scalar turbulence model}\label{sec: analysis-NL}

Given the fact that turbulent transport is a stochastic process, sophisticated statistical analysis of the mathematical models for such phenomena has been a center of attention in turbulence research. In this study, we consider the single- and two-point statistical quantities of interest in passive scalar transport to examine the performance of our nonlocal modeling within an equilibrium turbulent state against its conventional counterpart.

\subsection{Transport of the scalar variance}\label{subsec: ScVar}

\begin{figure}
    \begin{center}
        \begin{minipage}[b]{.69\linewidth}
            \centering
            \includegraphics[width=1\textwidth]{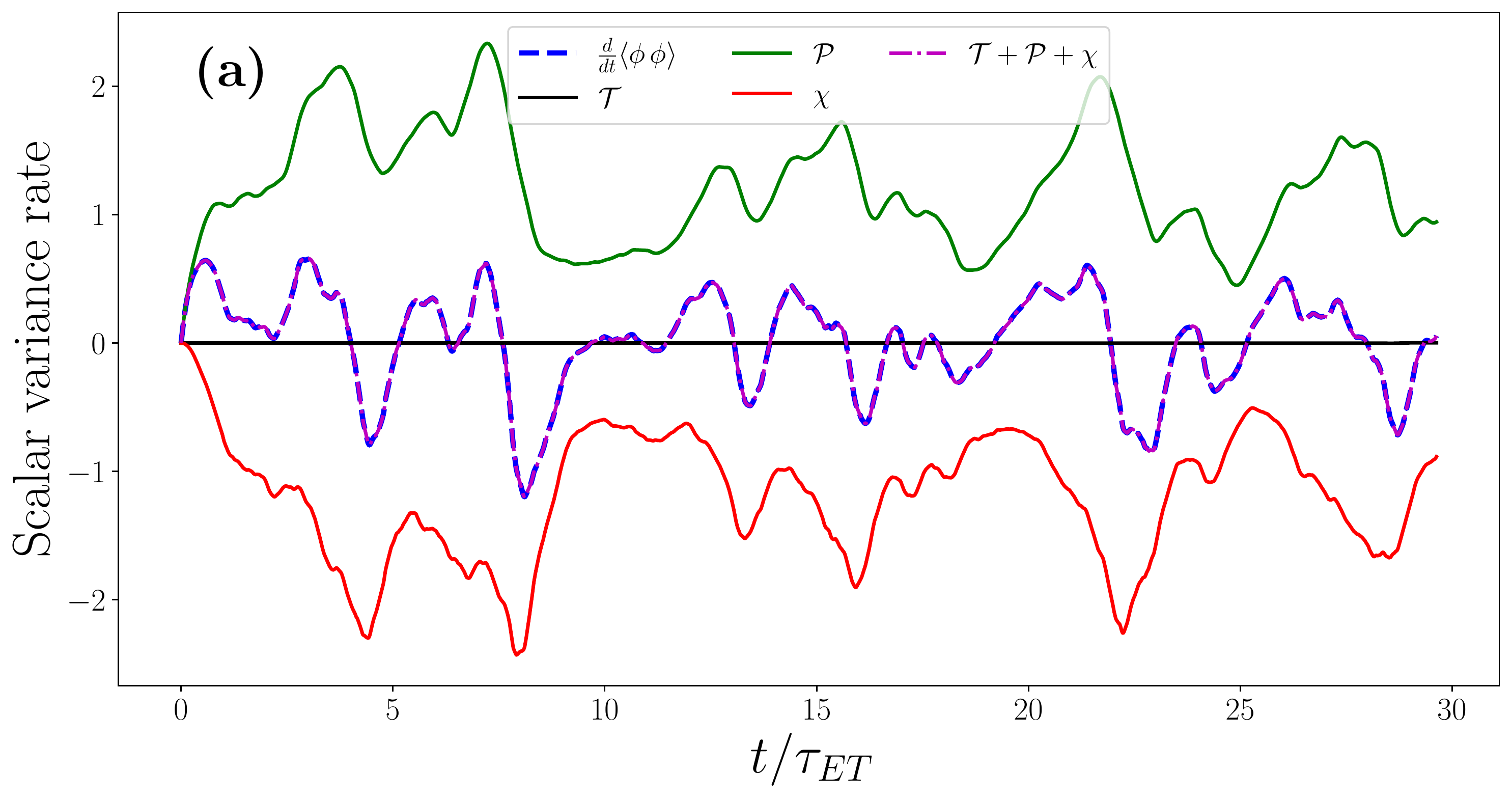}
        \end{minipage}

        \begin{minipage}[b]{.69\linewidth}
            \centering
            \includegraphics[width=1\textwidth]{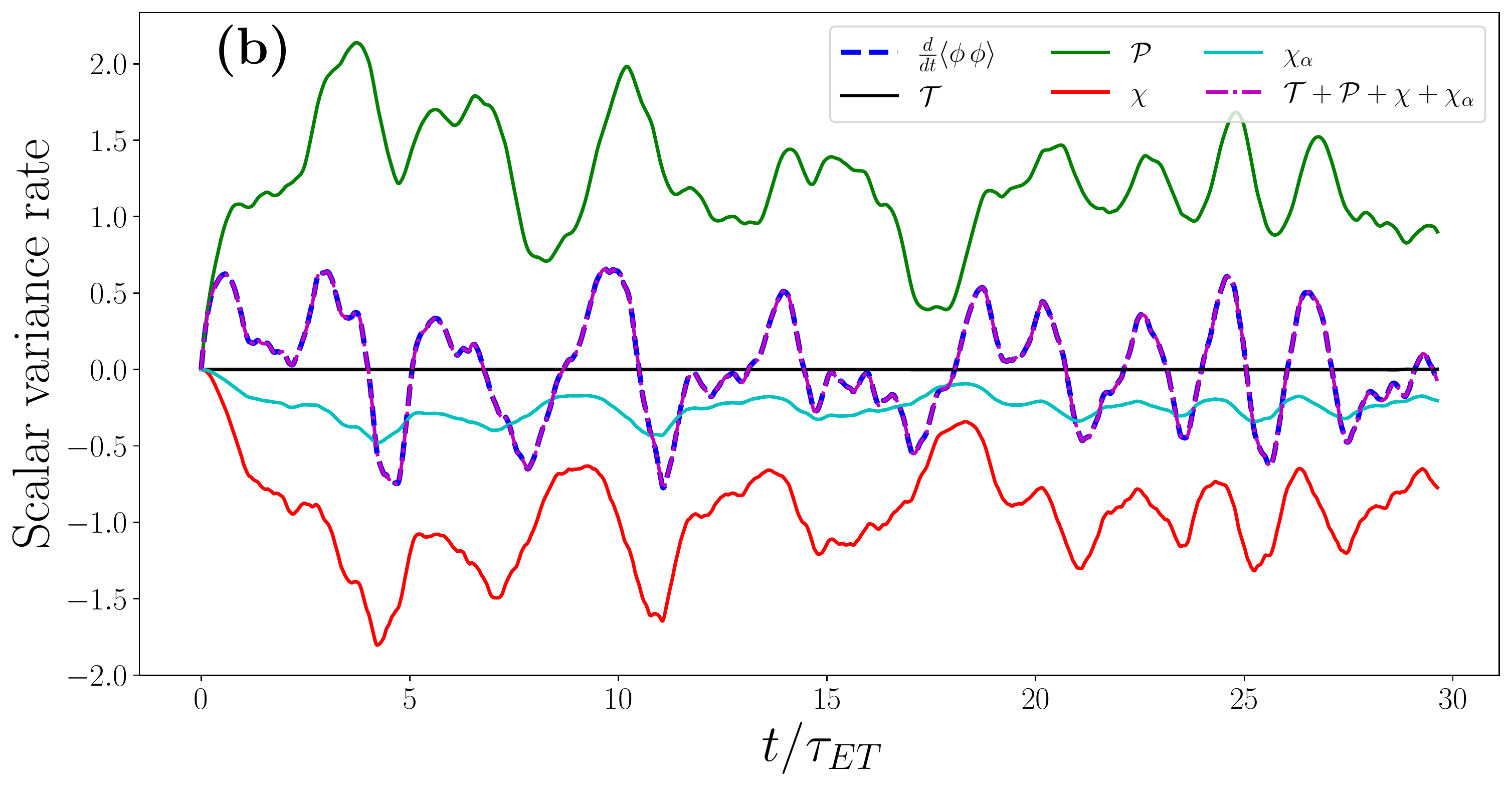}
        \end{minipage}
        \caption{Records of the contributing terms in the time-evolution of scalar variance, (a) standard model, (b) nonlocal fractional-order model. The time-averaged quantities are reported in Table \ref{tab: var_timeave}.}\label{fig: Balance_var}
    \end{center}
\end{figure}

\begin{figure}
    \begin{center}
        \includegraphics[width=0.8\textwidth]{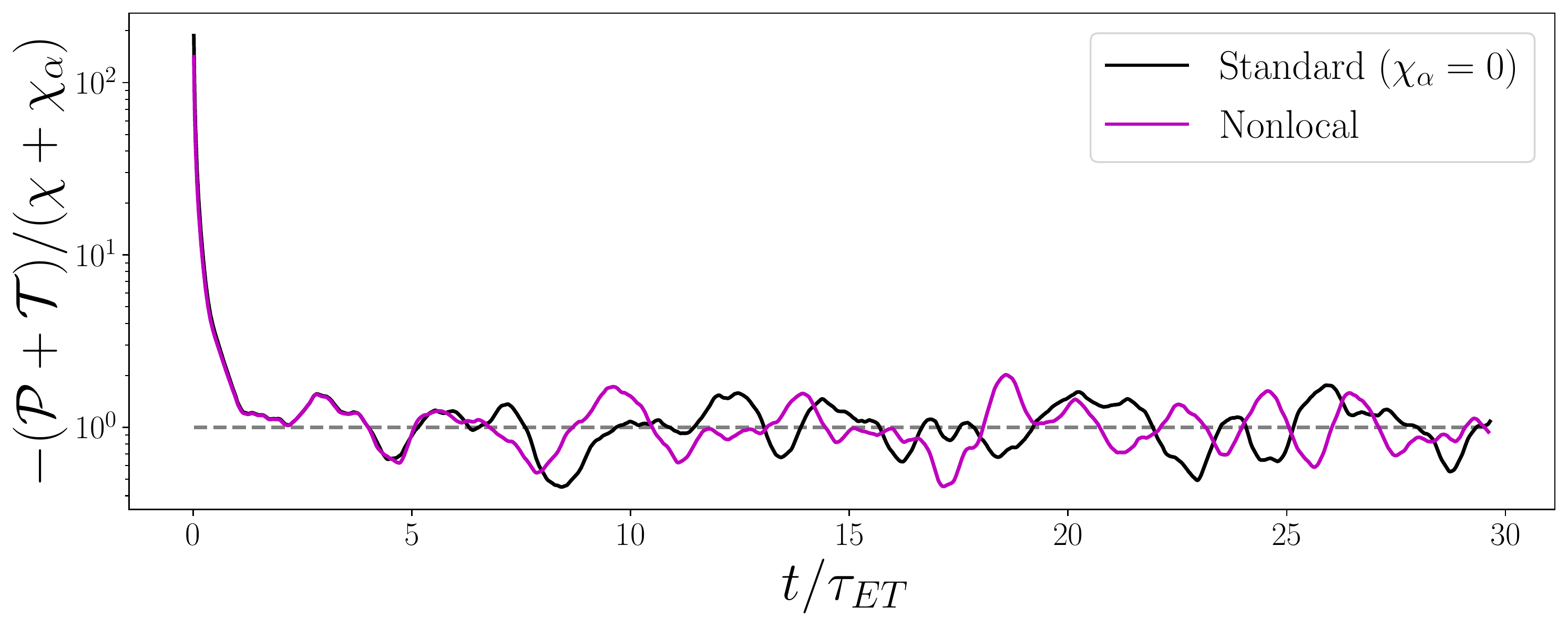}
        \caption{Tracking the record of the balance in scalar variance equation ensuring the equilibrium state in simulations of standard and nonlocal models.}\label{fig: Equilibrium}
        \end{center}
\end{figure}

\begin{table}
	\centering
	\begin{tabular}{ccccccccc}
			     &  $\qquad$ & $\mathcal{T}$  &  $\qquad$ & $\mathcal{P}$  &  $\qquad$ & $\chi$  &  $\qquad$ & $\chi_\alpha$ \\ \hline
		Standard   &  $\qquad$  & -0.00056  &  $\qquad$  & 1.1146  &  $\qquad$  & -1.1077 &  $\qquad$  & -- \\
		Nonlocal &  $\qquad$ & -0.00033  &  $\qquad$  & 1.1427  &  $\qquad$  & -0.9261 &  $\qquad$  & -0.2462 \\
	\end{tabular}
		\caption{Time-averaged values of the contributing terms in the time-evolution of scalar variance over the statistically stationary region.}\label{tab: var_timeave}
\end{table}

Transport of the scalar variance provides an important information about the evolution of the turbulence intensity. In fact, computational fluid dynamics approach make it possible to identify and keep track of the records of different influential mechanisms obtained from the mathematical modeling of the physics. It is well known that the multiplying both sides of the AD equation would yield the time-evolution equation for the turbulent intensity, $\phi^2$. Therefore, applying that to equation \eqref{eqn: std-AD-nl}, after using the incompressibility condition and chain rule for the spatial derivatives, one can derive that
\begin{align}\label{eqn: evl_scalar_var1}
    \frac{1}{2}\frac{\partial \phi^2}{\dt} =& - \left[ \nabla \cdot (\u \, \phi^2) - (\u \, \phi)\cdot \nabla \phi \right] - \boldsymbol{G} \cdot (\u\phi) \nonumber
    \\ & + \D \, \nabla \cdot (\phi \, \nabla \phi) - \D \, \nabla \phi \cdot \nabla \phi 
    \\ & + \D_\alpha \, \nabla \cdot \left( \phi \, \boldsymbol{\mathcal{R}}(-\Delta)^{\alpha-1/2}\phi \right) - \D_\alpha \, \nabla \phi \cdot \boldsymbol{\mathcal{R}}(-\Delta)^{\alpha-1/2}\phi, \nonumber
\end{align}
where $\boldsymbol{\mathcal{R}}(-\Delta)^{\alpha-1/2}(\cdot)$ denotes the fractional-order gradient obtained from the Riesz transform \citep{samiee2020fractional, akhavan2020data, samiee2021tempered}. Due to the homogeneity of the scalar fluctuations, averaging over the spatial domain is equivalent to the ensemble-averaging operation $\langle \, \cdot \, \rangle$ \citep{pope2001turbulent}. Thus, applying this averaging operation to \eqref{eqn: evl_scalar_var1} and considering that homogeneity of the fluctuating fields induces $\langle \nabla \cdot \, (\cdot) \rangle = \nabla \cdot \, (\langle \, \cdot \, \rangle) = 0$, the evolution of scalar variance $\langle \phi^2 \rangle$ is obtained as follows:
\begin{align}\label{eqn: evl_scalar_var2}
    \frac{1}{2}\frac{d}{dt} \langle\phi^2\rangle = \underbrace{\langle (\u\phi)\cdot \nabla \phi \rangle}_{\mathcal{T}} \underbrace{- \langle \boldsymbol{G} \cdot (\u\phi) \rangle}_{\mathcal{P}} - \underbrace{\D \, \langle \nabla \phi \cdot \nabla \phi \rangle}_{\chi} -\underbrace{\D_\alpha \, \langle \nabla \phi \cdot \boldsymbol{\mathcal{R}}(-\Delta)^{\alpha-1/2}\phi\rangle}_{\chi_\alpha} .
\end{align}
In \eqref{eqn: evl_scalar_var2}, the rate of scalar variance is composed of a balance between the turbulent advection effects ($\mathcal{T}$), production by the imposed mean-gradient ($\mathcal{P}$), molecular diffusion ($\chi$), and the nonlocal diffusion ($\chi_\alpha$). It is clear that for the standard scalar transport model in which $\D_\alpha = 0$, the nonlocal diffusion is consequently zero. According to the simulation considerations described in Section \ref{subsec: discretization}, we collect the records of the contributing terms in the right-hand side of \eqref{eqn: evl_scalar_var2}, and Figure \ref{fig: Balance_var} illustrates these time records for the standard and nonlocal models. Moreover, during both of the simulations we compute the rate of the scalar variance, $\frac{d}{dt}\langle \phi^2 \rangle$, using a forward-Euler finite difference scheme, and compare it with the record of the right-hand side of equation \eqref{eqn: evl_scalar_var2} constructed from the summation of the collected records. For both of the simulations an excellent match between these two computed quantities is observed during the entire simulation time as shown in Figure \ref{fig: Balance_var}. In the current work, we are focused to examine the statistical behavior of the developed nonlocal model at the ``turbulence equilibrium'' state. In this context, equilibrium is interpreted when the rate of scalar variance is considered when the following condition statistically holds:
\begin{align}\label{eqn: Equilibrium}
    \frac{\mathcal{P}+\mathcal{T}}{\chi+\chi_\alpha} \sim 1.
\end{align}
Figure \ref{fig: Equilibrium} shows the displays the record of this quantity for standard and nonlocal models and we notice that after approximately two large-eddy turnover times from resolving the scalar concentration, $(\mathcal{P}+\mathcal{T})/(\chi+\chi_\alpha)$ starts to fluctuate around 1. In order to make sure that the transient numerical effects are well past, we continue to simulate up to $t/\tau_{ET} = 10$, and consider the rest of simulation statistics in the fully developed turbulent equilibrium state. Therefore, the time-averaging operations in our study is performed on a sample space over $10 \leq t/\tau_{ET} \leq 30$. Accordingly, we can compute the time-averaged values of the contributing terms in evolution of scalar variance given in \eqref{eqn: evl_scalar_var2} as they are reported in Table \ref{tab: var_timeave}. Comparing the time-averaged values of $\mathcal{P}$ from both of the models reveals that the nonlocal model approximately includes 2.5\% more production rate of the scalar variance by the large-scale scalar mean-gradient. A reasonable interpretation for this observation is that once the nonlocal transfer of the scalar variance transfer in the cascade process is correctly modeled; therefore, this excessive 2.5\% production rate is captured at the equilibrium state for scalar turbulence. In other words, devising a nonlocal turbulent dissipation model ($\chi_\alpha$) in the scalar variance cascade mechanism would enable a balance in the equilibrium state so that the nonlocal effects in turbulent transport originating from large-scale ``anisotropy'' source are better captured throughout the DNS.

\begin{figure}
    \begin{center}
        \includegraphics[width=.7\textwidth]{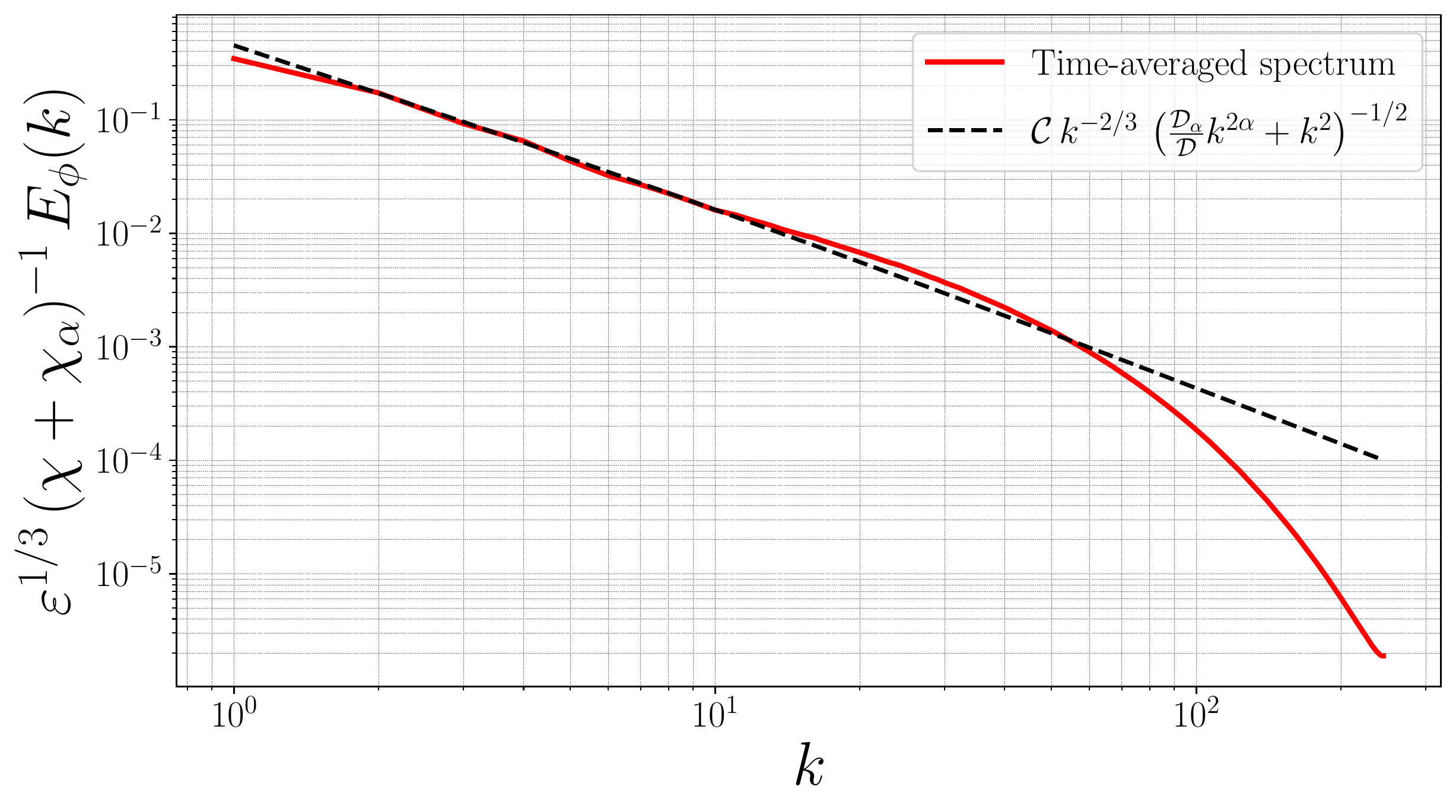}
        \caption{Time-averaged scalar spectrum computed from the data simulated with the nonlocal model, and evaluation of the identified scaling law in \eqref{eqn: Sc-sp-scaling2} for the scalar variance spectrum.}\label{fig: Spectra_sc_posteriori}
        \end{center}
\end{figure}

Finally, we compute the time-averaged scalar variance spectrum obtained from the scalar field resolved by the nonlocal fractional-order model to examine the modified scaling law \eqref{eqn: Sc-sp-scaling2} \textit{a posteriori}. Figure \ref{fig: Spectra_sc_posteriori} depicts this spectrum and reveals that the scaling \eqref{eqn: Sc-sp-scaling2} seamlessly holds. It is worth emphasizing that the total turbulent dissipation is denoted by $\chi+\chi_\alpha$.

\subsection{High-order small-scale statistics of scalar fluctuations}\label{subsec: SG_stats}

\begin{figure}
    \begin{center}
        \begin{minipage}[b]{.75\linewidth}
            \centering
            \includegraphics[width=1\textwidth]{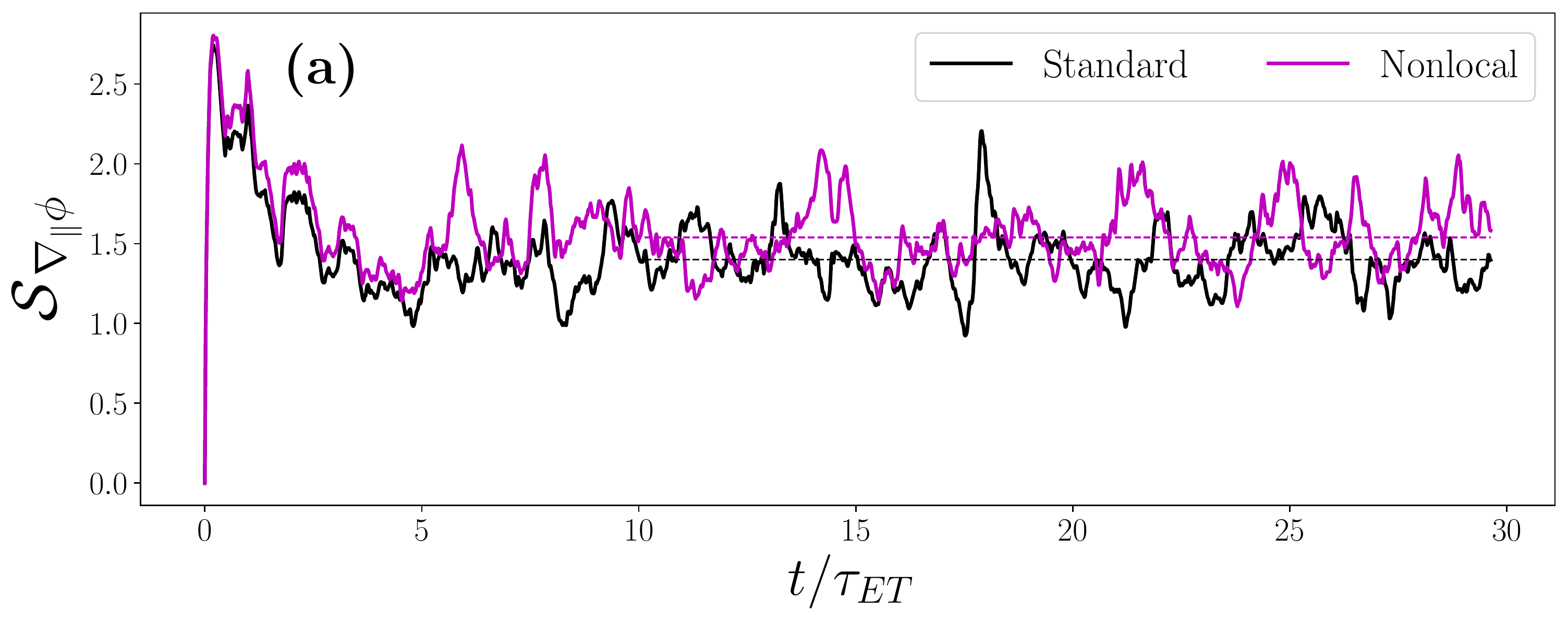}
        \end{minipage}
        
        \begin{minipage}[b]{.75\linewidth}
            \centering
            \includegraphics[width=1\textwidth]{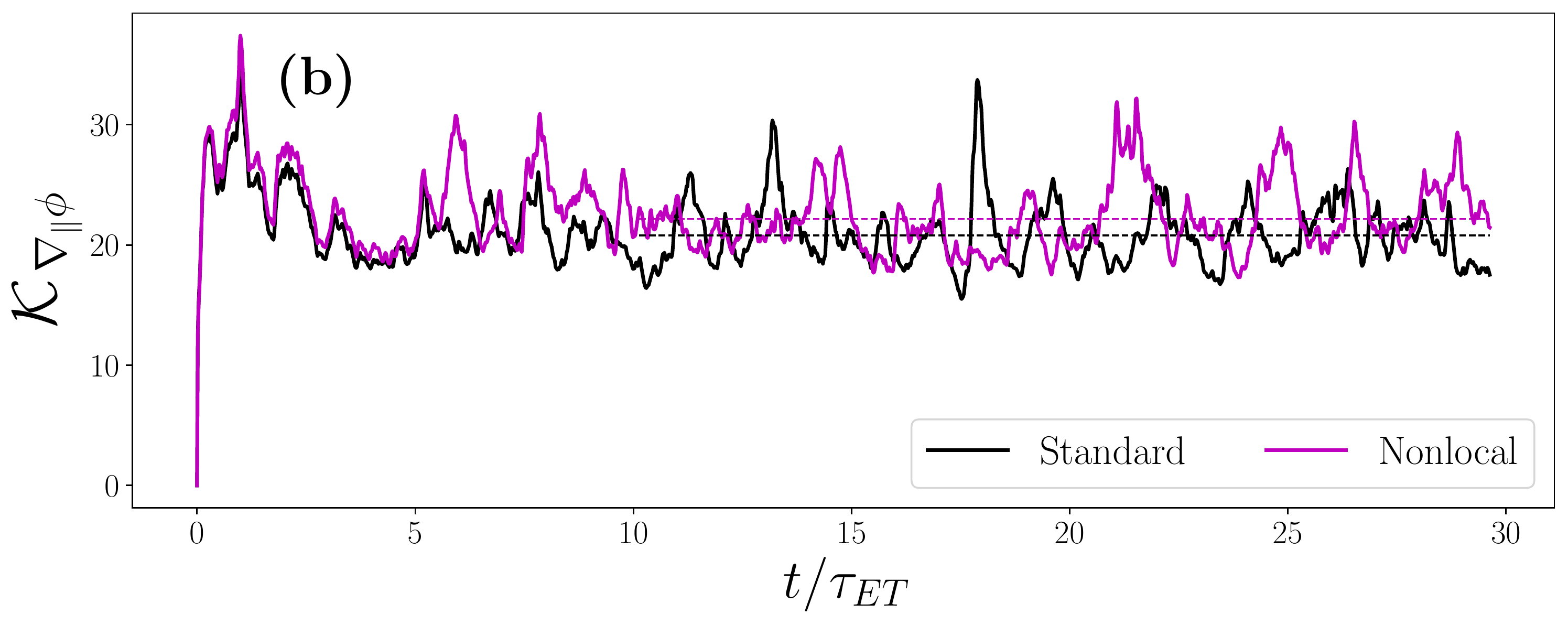}
        \end{minipage}
        \caption{Time records of (a) skewness, and (b) flatness of the scalar gradient along the anisotropy direction labeled by $\parallel$. The time-averaged values are identified with dashed lines over the statistically stationary state, and their values are reported in Table \ref{tab: moments_SG_timeave}.}\label{fig: moments_SG}
    \end{center}
\end{figure}

It is well-understood that statistics of the turbulence at the small-scales of the transport are represented through the central moments of the gradients of the fluctuating fields. Here, we are interested in discovering the small-scale statistics when the scalar field is sufficiently resolved with the proposed nonlocal scalar transport model. In fact, we seek to understand what would be the prediction of this model for the asymmetric and highly intermittent nature of passive scalar turbulence at the small scales. Therefore, we compute the skewness and flatness factors for the fluctuating concentration gradient, and due to the importance of the statistical behavior along the anisotropy direction $\parallel$, we focus on $\mathcal{S}_{\, \nabla_\parallel\phi}$ and $\mathcal{K}_{\, \nabla_\parallel\phi}$. Figure \ref{fig: moments_SG} illustrates the records of these two statistical quantities throughout the entire simulation for the standard and nonlocal models. Over the equilibrium state, explained in section \ref{subsec: ScVar}, we obtain the mean values of the of these statistical quantities by time-averaging, and their values are reported in Table \ref{tab: moments_SG_timeave}. These time-averaged values show that the nonlocal model yields the skewness factor 10\% more than the standard model, and the flatness factor is approximately 7\% higher in the nonlocal transport model. This evidently implies that an appropriate modeling of the nonlocal turbulent scalar transfer via the fractional-order model properly reflects the well-known statistical features of highly non-Gaussian behavior of the passive scalar turbulence reported in the literature \citep{warhaft2000passive, sreenivasan2019turbulent}.

\begin{table}
	\centering
	\begin{tabular}{ccccc}
			     &  $\qquad$ & $\mathcal{S}_{\, \nabla_\parallel\phi}$    &  $\qquad$ & $\mathcal{K}_{\, \nabla_\parallel\phi}$ \\
		Standard model  &  $\qquad$  & 1.40  &  $\qquad$  & 20.8 \\
		Nonlocal model  &  $\qquad$ & 1.54  &  $\qquad$  & 22.2\\
	\end{tabular}
		\caption{Time-averaged values of $\mathcal{S}_{\, \nabla_\parallel\phi}$, and $\mathcal{K}_{\, \nabla_\parallel\phi}$ over the statistically stationary state as illustrated in Figure \ref{fig: moments_SG}.}\label{tab: moments_SG_timeave}
\end{table}

 =============================================================================================
\subsection{Two-point statistics and structure functions}\label{subsec: TP_Strs}

\begin{figure}
    \begin{center}
        \begin{minipage}[b]{.49\linewidth}
            \centering
            \includegraphics[width=1\textwidth]{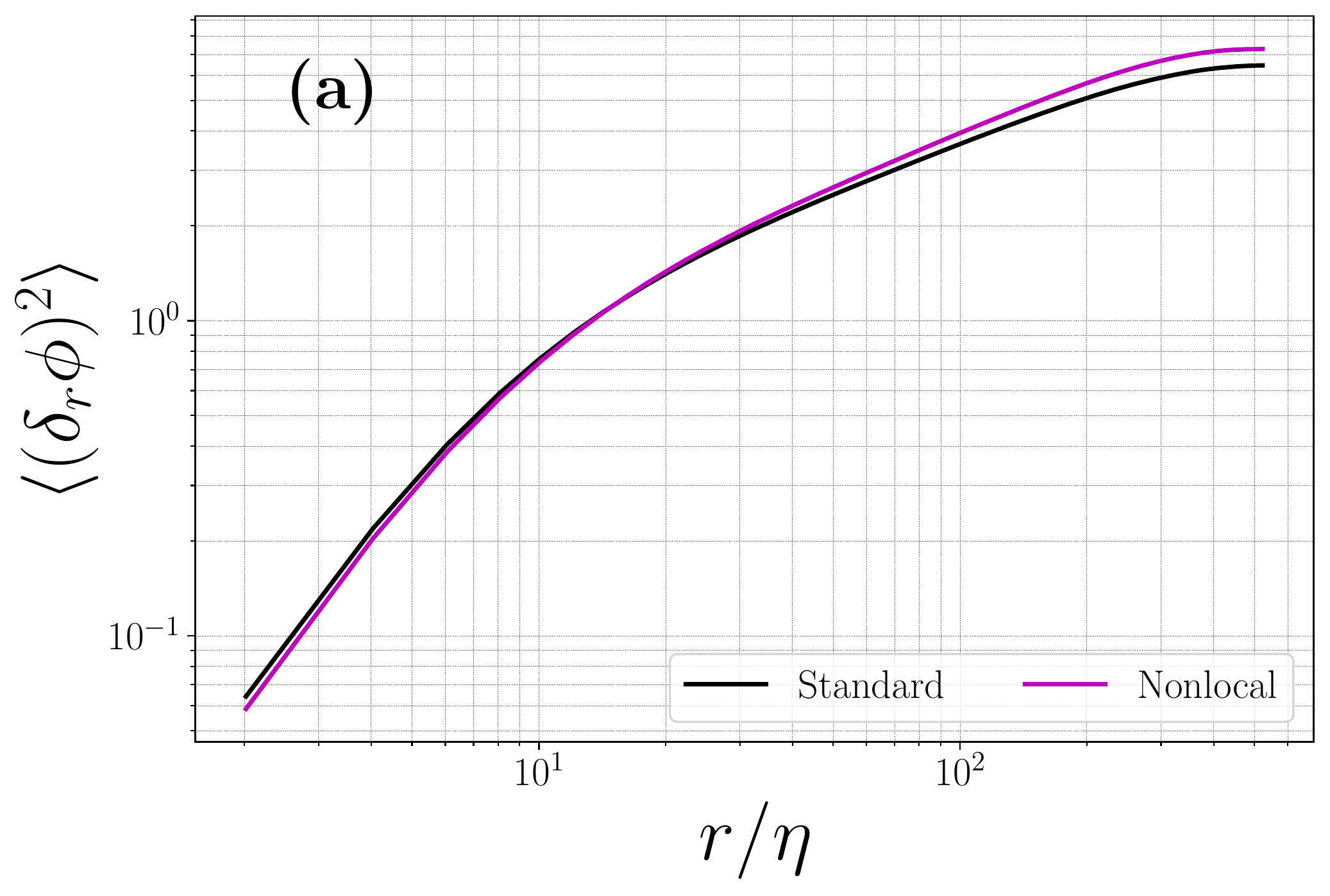}
        \end{minipage}
        \begin{minipage}[b]{.49\linewidth}
            \centering
            \includegraphics[width=1\textwidth]{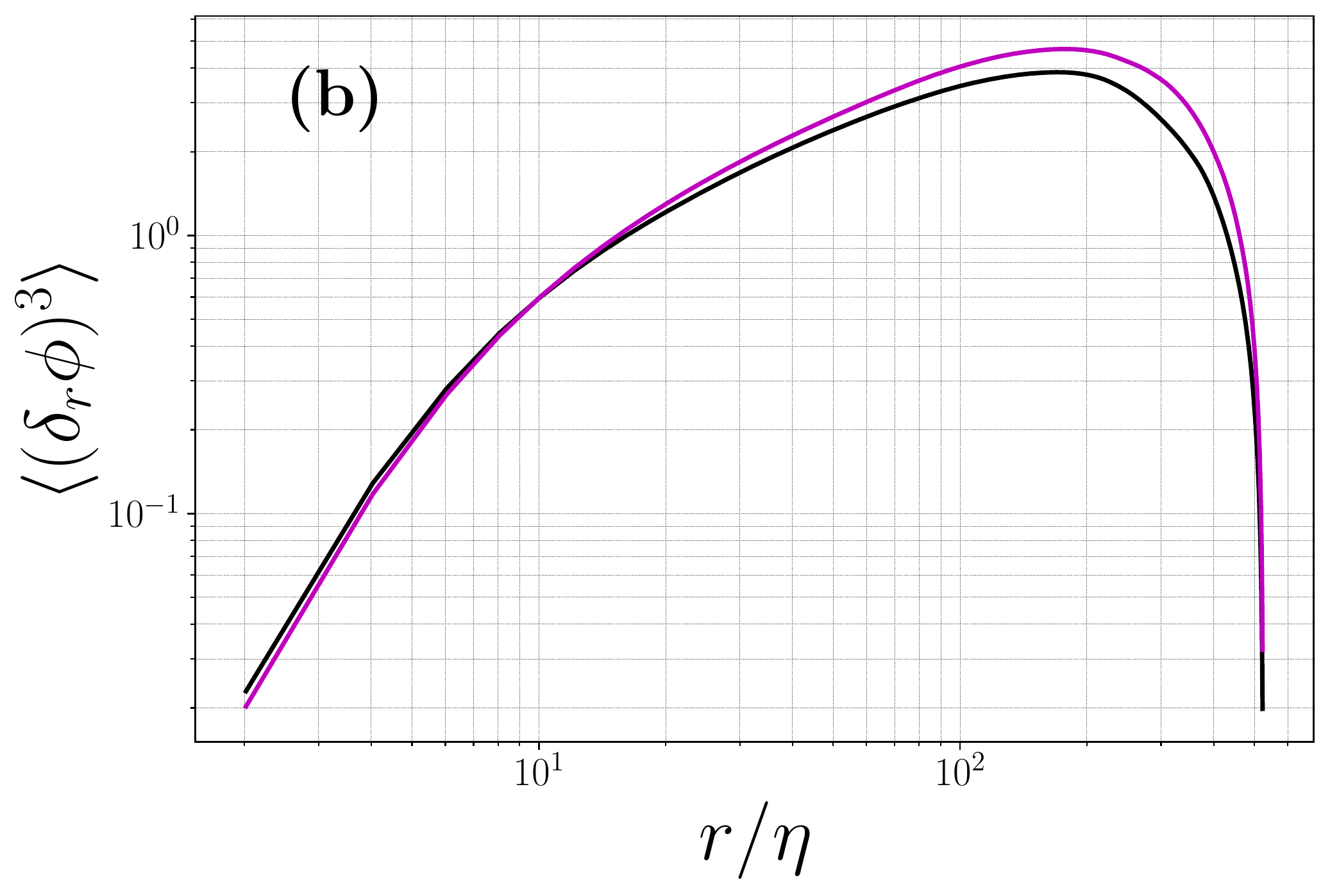}
        \end{minipage}
        \caption{Time-averaged $n$th-order scalar structure functions obtained from the simulations with standard and nonlocal model, with $r=2 \, \eta$. (a) $n=2$, and (b) $n=3$. }\label{fig: nth-order_strs}
    \end{center}
\end{figure}

Structure functions of order $n$ for a turbulent field such as scalar concentration are defined as: 
\begin{align}\label{eqn: str_funcs}
    	\langle (\delta_r \phi)^n \rangle = \langle [\phi(\boldsymbol{x}+\boldsymbol{r})-\phi(\boldsymbol{x})]^n \rangle, \quad n>1.
\end{align}
In \eqref{eqn: str_funcs}, $\boldsymbol{r}=r\boldsymbol{e}$ where $r$ is the increment of spatial shift, and $\boldsymbol{e}$ denotes a unit vector along a direction of interest. In fact, the structure functions would provide the $n$th-order statistics of spatial increments in the fluctuating field, which are interesting metrics in studying the nonlocality. In this study, we are interested in analyzing the behavior of the second- and third order structure function of $\phi$ along the direction of large-scale anisotropy, \textit{i.e.} $\boldsymbol{e}=(0,1,0)$, and regrading the size of the DNS grid, $r=2\,\eta$. Accordingly, Figure \ref{fig: nth-order_strs} shows the time-averaged (over the equilibrium turbulent region) structure functions of order 2 and 3 obtained from the simulations from standard and nonlocal scalar transport models. In Figure \ref{fig: nth-order_strs}(a), one can observe that for $r>40\,\eta$ the nonlocal second-order  structure function starts to exhibit higher values compared to the one computed from the simulations using standard model. For the third-order structure function values, the two models behave similarly up to $r/\eta = 10$; however, after that the nonlocal model shows higher values within the spatial shift domain associated with the inertial-convective and integral-scale domain. It is apparent that the maximum value of the time-averaged $\langle (\delta_r \phi)^3 \rangle$ in the nonlocal model is approximately 10 times higher than the standard model both occurring at $r/\eta \approx 200$.

\begin{figure}
    \begin{center}
        \includegraphics[width=.8\textwidth]{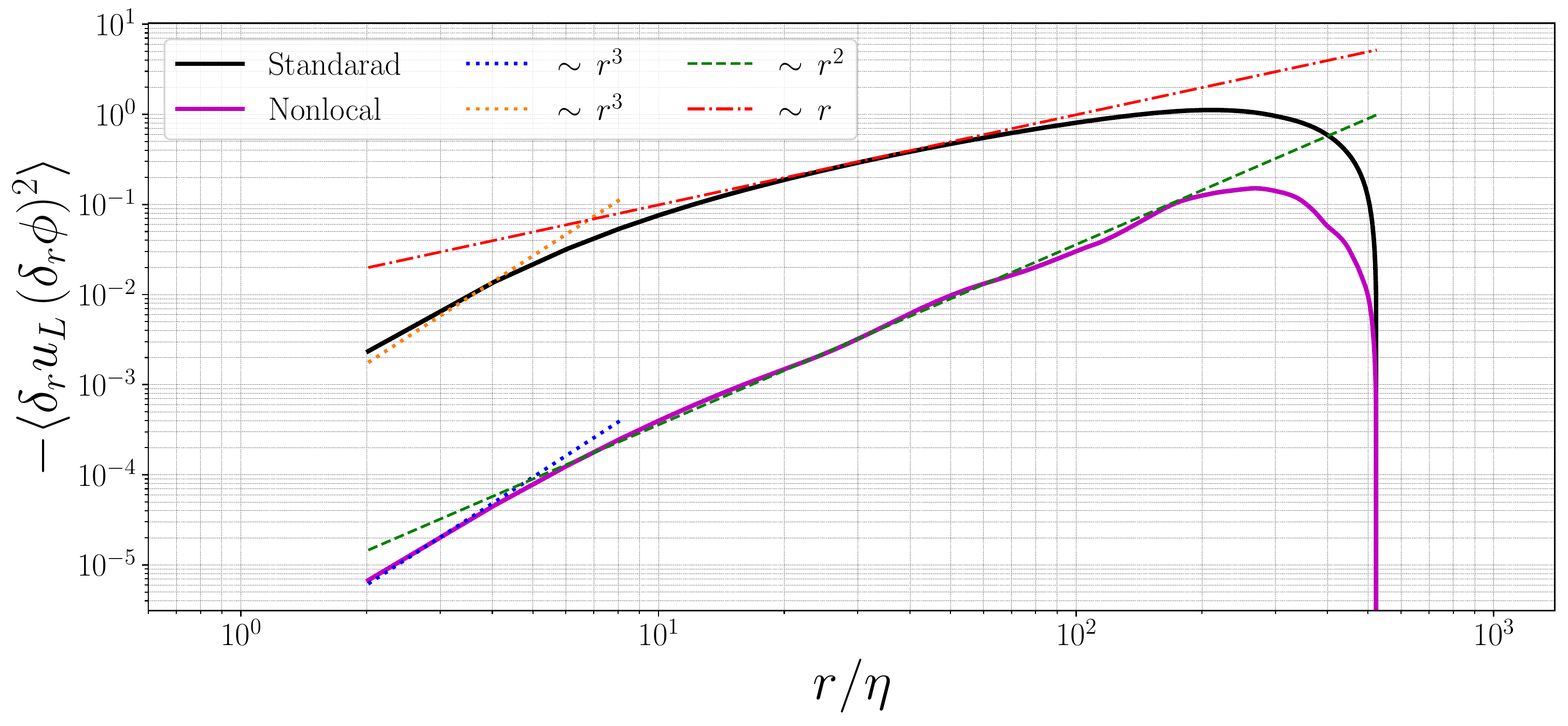}
        \caption{Third-order mixed longitudinal structure function, representing the statistics of advective increments. The nonlocal model shows a consistent and extended scaling over  universal range.}\label{fig: adv_3rd-order_str}
        \end{center}
\end{figure}

As initially introduced in \citep{yaglom1949local}, mixed ``velocity-scalar'' third-order structure function is an importasnt two-point statistical quantity measuring the advective turbulent transport in passive scalars. In the presence of large-scale anisotropy (imposed mean scalar gradient), the longitudinal contribution to this mixed structure function plays the dominant role in the advective transport \citep{iyer2014structure}, and its functional form obtained as $-\langle \delta_r u_L (\delta_r \phi)^2 \rangle$. Here, the subscript $L$ indicates the velocity component along the longitudinal direction with respect to spatial shift direction $\boldsymbol{r}$, where in our computational setup it would be $u_2$. Similar to the the second- and third-order scalar structure functions, we compute a time-averaged value for the $-\langle \delta_r u_L (\delta_r \phi)^2 \rangle$ over the stationary time domain. Figure \ref{fig: adv_3rd-order_str} shows that for the dissipative range ($r/\eta < 6$)  this structure function scales with $r^3$ in both standard and nonlocal models, while for almost the entire range of $6 < r/\eta < 200$ the mixed structure function obtained from the DNS with the nonlocal transport model scales with $r^2$. Unlike this universal-range scaling, one can observe that similar behavior is not necessarily seen in the $-\langle \delta_r u_L (\delta_r \phi)^2 \rangle$ when the scalar field is resolved with the standard model. However, in the standard model, a scaling with $r$ could be identified within $20 < r/\eta <60$. This comparison suggests that the considering the nonlocal effects in the turbulent cascade could result in emergence of more universal behavior in the two-point statistics of the advective transport, which inherently reveal high-order statistics of the nonlocality.

\section{Reconciliation with the fractional-order SGS modeling for LES}\label{sec: LES_rec}

\begin{figure}
    \begin{center}
        \includegraphics[width=.7\textwidth]{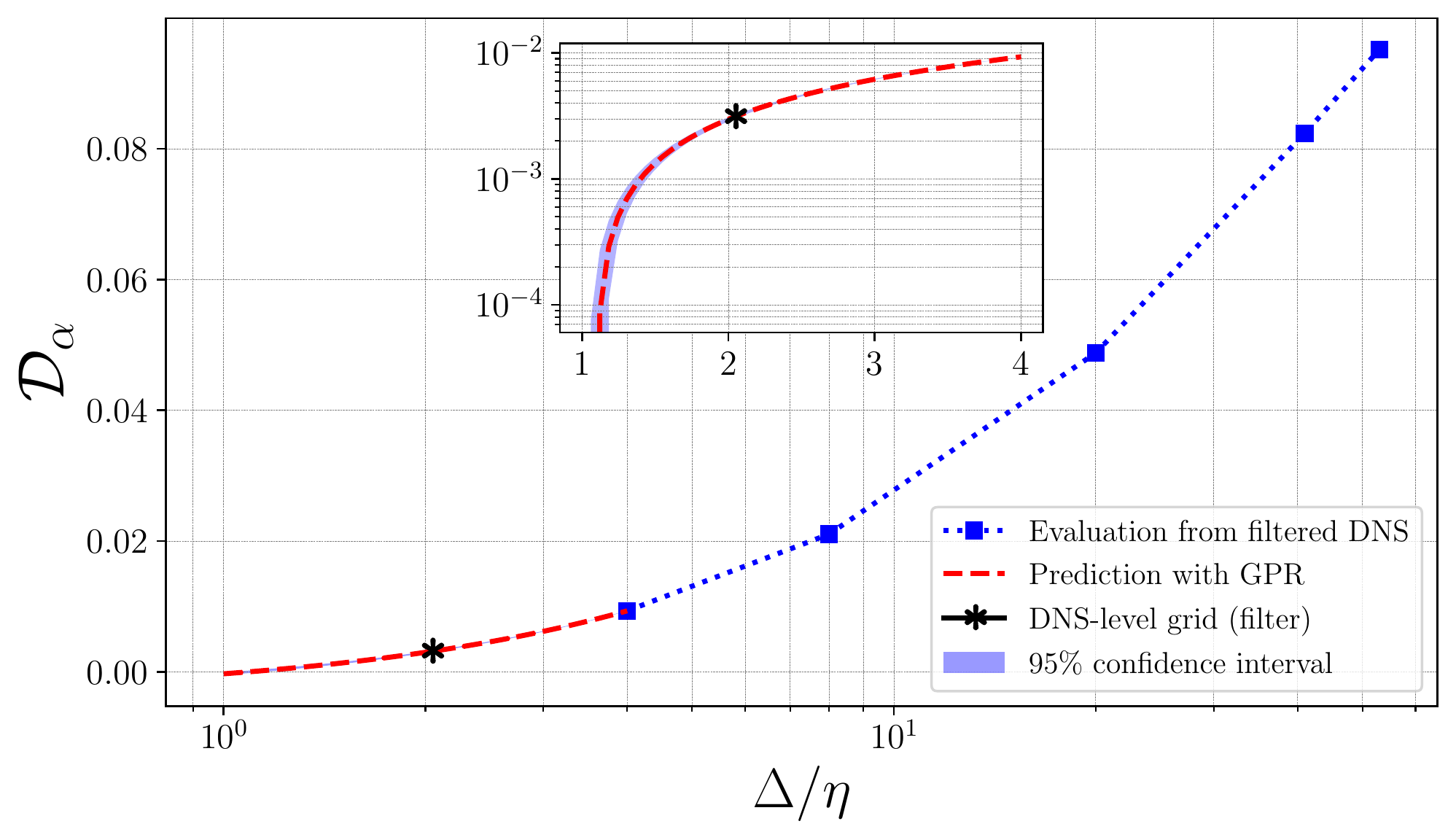}
        \caption{Reconciliation of the nonlocal model with the fractional-order SGS model developed in \citep{akhavan2020data} when the filter size is assumed to be at the dissipation range of $\Delta = 2 \, \eta$. The $\D_\alpha$ is computed from the filtered DNS data for $\Delta/\eta=4, 8, 20, 41, 52$ based on the data-driven methodology introduced in \citep{akhavan2020data} and a Gaussian process regressor (GPR) is trained based upon these evaluations, and $\D_\alpha$ is approximated based on the trained GPR for $\Delta/\eta < 4$ The predicted $\D_\alpha$ for $\Delta = 2 \, \eta$ is found to be in total agreement with the identified one obtained from the scaling analysis \textit{a priori} in Figure \ref{fig: nl-Sc-scalaing}.}\label{fig: LES_recons}
        \end{center}
\end{figure}

Recalling from section \ref{subsec: spectral-transfer-NL}, we performed an \textit{a priori} parameter identification that yielded $\alpha=0.65$ and $\D_\alpha/\D \approx 3.9$ (see Figure \ref{fig: nl-Sc-scalaing}). Here, our goal is to find a consistency between the currently developed model compared to the fractional-order SGS model introduced in \citep{akhavan2020data} when the filter scale is chosen close to the smallest scales of transport. In order to fulfill our goal, we need to show that given that the optimal fractional order for the SGS model $\alpha_{opt}=0.65$, what would be the value of $\D_\alpha/\D$ that is obtained from the procedure introduced in \citep{akhavan2020data} that relies on explicitly filtered data and sparse regression. Taking the filtered data from the simulation based on the standard scalar transport model \eqref{eqn: std-AD2} with the time-averaged scalar spectrum shown in Figure \ref{fig: nl-Sc-scalaing}, we can obtain the proportionality coefficient for the fractional-order SGS model. Here, we choose a top-hat box filtering kernel and obtain the filtered data for the filter sizes $\Delta/\eta=4, 8, 20, 41, 52$. Our goal is to evaluate $\D_\alpha$ when $\Delta/\eta = 2$; however, it is not computationally possible to obtain the filtered data for infer the $\D_\alpha$ at . Instead, we manage to employ a feasible machine learning algorithm (ML) to predict the desired $\D_\alpha$ while it is trained on the evaluated $\D_\alpha$ values from direct filtered data at larger filter sizes. Gaussian Process Regression is a known to be a suitable ML algorithm when one seeks to predict a quantity of interest from scarce experimental or computationally expensive high-fidelity data. Using the implementation of GPR in \texttt{Scikit-Learn} package \citep{scikit-learn}, we obtain the predicted value of $\D_\alpha/\D = 3.87$ for $\Delta = 2 \, \eta$ as illustrated in Figure \ref{fig: LES_recons}. This result shows that the \textit{a priori} estimates for the proportionality coefficient in the nonlocal scalar transport model is in great agreement with the fractional SGS model when filter size is selected as $\Delta = 2 \, \eta$; therefore, both models are reconciled. 

It is worth mentioning that the uncertainty in the predictions of the trained GPR for $\Delta/\eta < 4$ is assessed, and it is observed that the uncertainty is very low and practically negligible (see the 95\% confidence interval in Figure \ref{fig: LES_recons}).

\section{Conclusion and Remarks}\label{sec: Conclusion}

We proposed a modification to the spectral transfer model for the turbulent cascade of passive scalars under the effect of large-scale anisotropy. Employing the Corrsin's generalization to Onsager's turbulent cascade model, our modified model introduced an additional power-law term in the definition of local time-scale, $\tau(\k)$, in order to account for the induced nonlocal contributions originated from the anisotropy sources in the energy containing range. Subsequently, our approach yielded a modified scaling law for the passive scalar spectrum, $E_\phi(\k)$. This modified scaling showed a great match with the time-averaged 3-D scalar spectrum obtained from a well-resolved standard DNS after the parameter identification procedure. Using the integrated equation for the evolution of scalar spectrum, we revised the total scalar dissipation definition, which introduced an additional term into the total scalar dissipation representing the integrated effects of nonlocal turbulent dissipation cascade. This modification to the scalar dissipation returned that a fractional-order Laplacian acting on the scalar concentration is required in the AD equation. Using this revised AD equation, we performed a DNS study to analyze different quantities of turbulent transport compared to the DNS results obtained from convectional model at the statistical equilibrium state. Our analysis on the rate of the scalar variance showed that considering the effects of nonlocality in the scalar dissipation results in pronounced prediction of the production rate of scalar variance by the imposed mean-gradient (large-scale anisotropy), which could be interpreted in consistency with the breakdown of local isotropy in small scales of passive scalars. On the other hand, we showed that incorporation of the nonlocality effects in the scalar dissipation provides a more pronounced prediction for time-averaged records of the skewness and flatness factors for the $\nabla_{\Vert}\phi$, confirming the essence of devising a proper modeling mechanism for cascade of the anisotropy effects from large to small scales. Moreover, a two-point statistical analysis for the advective scalar increments (the third-order mixed longitudinal structure function) revealed that the DNS results obtained from nonlocal model provides a long-range scaling with $r^2$. This observation on long-range scaling suggested that the inclusion of nonlocal cascading mechanism in the presence of large-scale anisotropy could result in prediction of more universal behavior over a wide span of scales in turbulent scalar transport. Finally, we showed an accurate consistency between the developed spectral transfer model and the fractional-order SGS modeling with $\Delta \approx 2\, \eta$, after employing a well-trained GPR model.

\section*{Acknowledgement}

This work was supported by the ARO YIP award (W911NF-19-1-0444), and partially by the MURI/ARO grant (W911NF-15-1-0562) and the NSF award (DMS-1923201). The HPC resources and services were provided by the Institute for Cyber-Enabled Research (ICER) at Michigan State University.

\bibliographystyle{jfm}
\bibliography{mybib}

\begin{thebibliography}{63}
\expandafter\ifx\csname natexlab\endcsname\relax\def\natexlab#1{#1}\fi
\def\au#1{#1} \def\ed#1{#1} \def\yr#1{#1}\def\at#1{#1}\def\jt#1{\textit{#1}}
  \def\bt#1{#1}\def\bvol#1{\textbf{#1}} \def\vol#1{#1} \def\pg#1{#1}
  \def\publ#1{#1}\def\arxiv#1{#1}\def\org#1{#1}\def\st#1{\textit{#1}}

\bibitem[Akhavan-Safaei {\em et~al.\/}(2021)Akhavan-Safaei, Samiee \&
  Zayernouri]{akhavan2020data}
{\sc \au{Akhavan-Safaei, Ali}, \au{Samiee, Mehdi} \& \au{Zayernouri, Mohsen}}
  \yr{2021}  \at{{Data-driven fractional subgrid-scale modeling for scalar
  turbulence: A nonlocal LES approach}}.  \jt{Journal of Computational Physics}
   \bvol{446},  \pg{110571}.

\bibitem[Akhavan-Safaei \& Zayernouri(2020)]{akhavan2020parallel}
{\sc \au{Akhavan-Safaei, Ali} \& \au{Zayernouri, Mohsen}} \yr{2020}  \at{{A
  Parallel Integrated Computational-Statistical Platform for Turbulent
  Transport Phenomena}}.  \jt{arXiv preprint arXiv:2012.04838} .

\bibitem[Batchelor {\em et~al.\/}(1959)Batchelor, Howells \&
  Townsend]{batchelor1959b}
{\sc \au{Batchelor, GK}, \au{Howells, ID} \& \au{Townsend, AA}} \yr{1959}
  \at{{Small-scale variation of convected quantities like temperature in
  turbulent fluid Part 2. The case of large conductivity}}.  \jt{Journal of
  Fluid Mechanics}  \bvol{5}~(1),  \pg{134--139}.

\bibitem[Batchelor(1959)]{batchelor1959a}
{\sc \au{Batchelor, George~K}} \yr{1959}  \at{{Small-scale variation of
  convected quantities like temperature in turbulent fluid Part 1. General
  discussion and the case of small conductivity}}.  \jt{Journal of Fluid
  Mechanics}  \bvol{5}~(1),  \pg{113--133}.

\bibitem[Buaria {\em et~al.\/}(2021)Buaria, Clay, Sreenivasan \&
  Yeung]{buaria2021small}
{\sc \au{Buaria, Dhawal}, \au{Clay, Matthew~P}, \au{Sreenivasan, Katepalli~R}
  \& \au{Yeung, PK}} \yr{2021}  \at{Small-scale isotropy and ramp-cliff
  structures in scalar turbulence}.  \jt{Physical Review Letters}
  \bvol{126}~(3),  \pg{034504}.

\bibitem[Chertkov {\em et~al.\/}(1995)Chertkov, Falkovich, Kolokolov \&
  Lebedev]{chertkov1995normal}
{\sc \au{Chertkov, M}, \au{Falkovich, Gregory}, \au{Kolokolov, I} \&
  \au{Lebedev, V}} \yr{1995}  \at{Normal and anomalous scaling of the
  fourth-order correlation function of a randomly advected passive scalar}.
  \jt{Physical Review E}  \bvol{52}~(5),  \pg{4924}.

\bibitem[Chertkov {\em et~al.\/}(1999)Chertkov, Pumir \&
  Shraiman]{chertkov1999lagrangian}
{\sc \au{Chertkov, Michael}, \au{Pumir, Alain} \& \au{Shraiman, Boris~I}}
  \yr{1999}  \at{Lagrangian tetrad dynamics and the phenomenology of
  turbulence}.  \jt{Physics of fluids}  \bvol{11}~(8),  \pg{2394--2410}.

\bibitem[Cheung \& Zaki(2014)]{cheung2014exact}
{\sc \au{Cheung, Lawrence~C} \& \au{Zaki, Tamer~A}} \yr{2014}  \at{An exact
  representation of the nonlinear triad interaction terms in spectral space}.
  \jt{Journal of fluid mechanics}  \bvol{748},  \pg{175--188}.

\bibitem[Corrsin(1951)]{corrsin1951spectrum}
{\sc \au{Corrsin, Stanley}} \yr{1951}  \at{On the spectrum of isotropic
  temperature fluctuations in an isotropic turbulence}.  \jt{Journal of Applied
  Physics}  \bvol{22}~(4),  \pg{469--473}.

\bibitem[Corrsin(1964)]{Corrsin1964PoF}
{\sc \au{Corrsin, Stanley}} \yr{1964}  \at{{Further Generalization of Onsager's
  Cascade Model for Turbulent Spectra}}.  \jt{The Physics of Fluids}
  \bvol{7}~(8),  \pg{1156--1159}.

\bibitem[Di~Leoni {\em et~al.\/}(2021)Di~Leoni, Zaki, Karniadakis \&
  Meneveau]{di2021two}
{\sc \au{Di~Leoni, Patricio~Clark}, \au{Zaki, Tamer~A}, \au{Karniadakis,
  George} \& \au{Meneveau, Charles}} \yr{2021}  \at{Two-point
  stress--strain-rate correlation structure and non-local eddy viscosity in
  turbulent flows}.  \jt{Journal of Fluid Mechanics}  \bvol{914}.

\bibitem[Dimotakis(2005)]{dimotakis2005turbulent}
{\sc \au{Dimotakis, Paul~E}} \yr{2005}  \at{Turbulent mixing}.  \jt{Annu. Rev.
  Fluid Mech.}  \bvol{37},  \pg{329--356}.

\bibitem[Domaradzki \& Rogallo(1990)]{domaradzki1990local}
{\sc \au{Domaradzki, J~Andrzej} \& \au{Rogallo, Robert~S}} \yr{1990}  \at{Local
  energy transfer and nonlocal interactions in homogeneous, isotropic
  turbulence}.  \jt{Physics of Fluids A: Fluid Dynamics}  \bvol{2}~(3),
  \pg{413--426}.

\bibitem[Donzis \& Yeung(2010)]{donzis2010resolution}
{\sc \au{Donzis, DA} \& \au{Yeung, PK}} \yr{2010}  \at{Resolution effects and
  scaling in numerical simulations of passive scalar mixing in turbulence}.
  \jt{Physica D: Nonlinear Phenomena}  \bvol{239}~(14),  \pg{1278--1287}.

\bibitem[Donzis {\em et~al.\/}(2008)Donzis, Yeung \&
  Sreenivasan]{donzis2008dissipation}
{\sc \au{Donzis, DA}, \au{Yeung, PK} \& \au{Sreenivasan, KR}} \yr{2008}
  \at{Dissipation and enstrophy in isotropic turbulence: resolution effects and
  scaling in direct numerical simulations}.  \jt{Physics of Fluids}
  \bvol{20}~(4),  \pg{045108}.

\bibitem[Failla \& Zingales(2020)]{failla2020advanced}
{\sc \au{Failla, Giuseppe} \& \au{Zingales, Massimiliano}} \yr{2020} Advanced
  materials modelling via fractional calculus: challenges and perspectives.

\bibitem[Frisch {\em et~al.\/}(1998)Frisch, Mazzino \& Vergassola]{frisch1998}
{\sc \au{Frisch, U.}, \au{Mazzino, A.} \& \au{Vergassola, M.}} \yr{1998}
  \at{{Intermittency in Passive Scalar Advection}}.  \jt{Phys. Rev. Lett.}
  \bvol{80},  \pg{5532--5535}.

\bibitem[Gat {\em et~al.\/}(1998)Gat, Procaccia \& Zeitak]{gat1998}
{\sc \au{Gat, Omri}, \au{Procaccia, Itamar} \& \au{Zeitak, Reuven}} \yr{1998}
  \at{{Anomalous Scaling in Passive Scalar Advection: Monte Carlo Lagrangian
  Trajectories}}.  \jt{Phys. Rev. Lett.}  \bvol{80},  \pg{5536--5539}.

\bibitem[Gollub {\em et~al.\/}(1991)Gollub, Clarke, Gharib, Lane \&
  Mesquita]{gollub_1991}
{\sc \au{Gollub, J.~P.}, \au{Clarke, J.}, \au{Gharib, M.}, \au{Lane, B.} \&
  \au{Mesquita, O.~N.}} \yr{1991}  \at{Fluctuations and transport in a stirred
  fluid with a mean gradient}.  \jt{Phys. Rev. Lett.}  \bvol{67},
  \pg{3507--3510}.

\bibitem[Hill(1978)]{hill1978models}
{\sc \au{Hill, Reginald~J}} \yr{1978}  \at{Models of the scalar spectrum for
  turbulent advection}.  \jt{Journal of Fluid Mechanics}  \bvol{88}~(3),
  \pg{541--562}.

\bibitem[Iyer \& Yeung(2014)]{iyer2014structure}
{\sc \au{Iyer, KP} \& \au{Yeung, PK}} \yr{2014}  \at{{Structure functions and
  applicability of Yaglom's relation in passive-scalar turbulent mixing at low
  Schmidt numbers with uniform mean gradient}}.  \jt{Physics of Fluids}
  \bvol{26}~(8),  \pg{085107}.

\bibitem[Jayesh \& Warhaft(1991)]{jayesh_1991}
{\sc \au{Jayesh} \& \au{Warhaft, Z.}} \yr{1991}  \at{Probability distribution
  of a passive scalar in grid-generated turbulence}.  \jt{Phys. Rev. Lett.}
  \bvol{67},  \pg{3503--3506}.

\bibitem[Jayesh \& Warhaft(1992)]{jayesh_1992}
{\sc \au{Jayesh} \& \au{Warhaft, Z.}} \yr{1992}  \at{Probability distribution,
  conditional dissipation, and transport of passive temperature fluctuations in
  grid‐generated turbulence}.  \jt{Physics of Fluids A: Fluid Dynamics}
  \bvol{4}~(10),  \pg{2292--2307}.

\bibitem[Kang \& Meneveau(2001)]{kang2001passive}
{\sc \au{Kang, Hyung~Suk} \& \au{Meneveau, Charles}} \yr{2001}  \at{Passive
  scalar anisotropy in a heated turbulent wake: new observations and
  implications for large-eddy simulations}.  \jt{Journal of Fluid Mechanics}
  \bvol{442},  \pg{161--170}.

\bibitem[Keith {\em et~al.\/}(2021)Keith, Khristenko \&
  Wohlmuth]{keith2021fractional}
{\sc \au{Keith, Brendan}, \au{Khristenko, Ustim} \& \au{Wohlmuth, Barbara}}
  \yr{2021}  \at{{A fractional PDE model for turbulent velocity fields near
  solid walls}}.  \jt{Journal of Fluid Mechanics}  \bvol{916}.

\bibitem[Kitamura(2021)]{kitamura2021spectral}
{\sc \au{Kitamura, Takuya}} \yr{2021}  \at{Spectral theory of passive scalar
  with mean scalar gradient}.  \jt{Journal of Fluid Mechanics}  \bvol{923}.

\bibitem[Kolmogorov(1941{\natexlab{{\em a\/}}})]{kolmogorov1941energy}
{\sc \au{Kolmogorov, A.~N.}} \yr{1941{\natexlab{{\em a\/}}}} Energy dissipation
  in locally isotropic turbulence.  \bt{In {\em Dokl. Akad. Nauk. SSSR\/}}, ,
  \vol{vol.~32},  \pg{pp. 19--21}.

\bibitem[Kolmogorov(1941{\natexlab{{\em b\/}}})]{kolmogorov1941local}
{\sc \au{Kolmogorov, A.~N.}} \yr{1941{\natexlab{{\em b\/}}}}  \at{Local
  structure of turbulence in incompressible fluid under very high values of
  reynolds numbers}.  \jt{Reports of AS USSR}  \bvol{30}~(4),  \pg{299--303}.

\bibitem[Kolmogorov(1962)]{kolmogorov1962refinement}
{\sc \au{Kolmogorov, Andrey~Nikolaevich}} \yr{1962}  \at{A refinement of
  previous hypotheses concerning the local structure of turbulence in a viscous
  incompressible fluid at high reynolds number}.  \jt{Journal of Fluid
  Mechanics}  \bvol{13}~(1),  \pg{82--85}.

\bibitem[Lane {\em et~al.\/}(1993)Lane, Mesquita, Meyers \& Gollub]{lane_1993}
{\sc \au{Lane, B.~R.}, \au{Mesquita, O.~N.}, \au{Meyers, S.~R.} \& \au{Gollub,
  J.~P.}} \yr{1993}  \at{Probability distributions and thermal transport in a
  turbulent grid flow}.  \jt{Physics of Fluids A: Fluid Dynamics}
  \bvol{5}~(9),  \pg{2255--2263}.

\bibitem[Li \& Meneveau(2006)]{li2006intermittency}
{\sc \au{Li, YI} \& \au{Meneveau, Charles}} \yr{2006}  \at{{Intermittency
  trends and Lagrangian evolution of non-Gaussian statistics in turbulent flow
  and scalar transport}}.  \jt{Journal of Fluid Mechanics}  \bvol{558},
  \pg{133--142}.

\bibitem[Mehta {\em et~al.\/}(2019)Mehta, Pang, Song \&
  Karniadakis]{mehta2019discovering}
{\sc \au{Mehta, Pavan~Pranjivan}, \au{Pang, Guofei}, \au{Song, Fangying} \&
  \au{Karniadakis, George~Em}} \yr{2019}  \at{Discovering a universal
  variable-order fractional model for turbulent couette flow using a
  physics-informed neural network}.  \jt{Fractional Calculus and Applied
  Analysis}  \bvol{22}~(6),  \pg{1675--1688}.

\bibitem[Meneveau \& Katz(2000)]{meneveau2000scale}
{\sc \au{Meneveau, Charles} \& \au{Katz, Joseph}} \yr{2000}
  \at{Scale-invariance and turbulence models for large-eddy simulation}.
  \jt{Annual Review of Fluid Mechanics}  \bvol{32}~(1),  \pg{1--32}.

\bibitem[Monin \& Yaglom(1975)]{monin_statistical}
{\sc \au{Monin, A.~S.} \& \au{Yaglom, A.~M.}} \yr{1975} {\em {Statistical fluid
  mechanics, volume II: mechanics of turbulence}\/}, ,  \vol{vol.~2}.
  \publ{MIT press}.

\bibitem[Moser {\em et~al.\/}(2021)Moser, Haering \&
  Yalla]{moser2020statistical}
{\sc \au{Moser, Robert~D}, \au{Haering, Sigfried~W} \& \au{Yalla, Gopal~R}}
  \yr{2021}  \at{Statistical properties of subgrid-scale turbulence models}.
  \jt{Annual Review of Fluid Mechanics}  \bvol{53}.

\bibitem[Oboukhov(1962)]{oboukhov_1962}
{\sc \au{Oboukhov, A.~M.}} \yr{1962}  \at{Some specific features of atmospheric
  tubulence}.  \jt{Journal of Fluid Mechanics}  \bvol{13}~(1),  \pg{77–81}.

\bibitem[Obukhov(1949)]{obukhov1949structure}
{\sc \au{Obukhov, A.~M.}} \yr{1949} Structure of temperature field in turbulent
  flow.  \bt{In {\em Izv. Akad. Nauk. SSSR, Georg. i Geofiz.\/}}, ,
  \vol{vol.~13},  \pg{pp. 58--69}.

\bibitem[Onsager(1945)]{onsager1945distribution}
{\sc \au{Onsager, L.}} \yr{1945} The distribution of energy in turbulence.
  \bt{In {\em Physical Review\/}}, ,  \vol{vol.~68},  \pg{pp. 286--286}.

\bibitem[{Onsager}(1949)]{onsager1949}
{\sc \au{{Onsager}, L.}} \yr{1949}  \at{{Statistical hydrodynamics}}.  \jt{Il
  Nuovo Cimento}  \bvol{6}~(2),  \pg{279--287}.

\bibitem[Overholt \& Pope(1996)]{overholt1996direct}
{\sc \au{Overholt, MR} \& \au{Pope, SB}} \yr{1996}  \at{Direct numerical
  simulation of a passive scalar with imposed mean gradient in isotropic
  turbulence}.  \jt{Physics of Fluids}  \bvol{8}~(11),  \pg{3128--3148}.

\bibitem[Pedregosa {\em et~al.\/}(2011)Pedregosa, Varoquaux, Gramfort, Michel,
  Thirion, Grisel, Blondel, Prettenhofer, Weiss, Dubourg, Vanderplas, Passos,
  Cournapeau, Brucher, Perrot \& Duchesnay]{scikit-learn}
{\sc \au{Pedregosa, F.}, \au{Varoquaux, G.}, \au{Gramfort, A.}, \au{Michel,
  V.}, \au{Thirion, B.}, \au{Grisel, O.}, \au{Blondel, M.}, \au{Prettenhofer,
  P.}, \au{Weiss, R.}, \au{Dubourg, V.}, \au{Vanderplas, J.}, \au{Passos, A.},
  \au{Cournapeau, D.}, \au{Brucher, M.}, \au{Perrot, M.} \& \au{Duchesnay, E.}}
  \yr{2011}  \at{{Scikit-learn: Machine Learning in Python}}.  \jt{Journal of
  Machine Learning Research}  \bvol{12},  \pg{2825--2830}.

\bibitem[Pope(2001)]{pope2001turbulent}
{\sc \au{Pope, Stephen~B}} \yr{2001} Turbulent flows.

\bibitem[Prasad {\em et~al.\/}(1988)Prasad, Meneveau \&
  Sreenivasan]{prasad1988multifractal}
{\sc \au{Prasad, RR}, \au{Meneveau, C} \& \au{Sreenivasan, KR}} \yr{1988}
  \at{Multifractal nature of the dissipation field of passive scalars in fully
  turbulent flows}.  \jt{Physical Review Letters}  \bvol{61}~(1),  \pg{74}.

\bibitem[Pumir \& Shraiman(1995)]{pumir1995}
{\sc \au{Pumir, Alain} \& \au{Shraiman, Boris~I.}} \yr{1995}  \at{{Persistent
  Small Scale Anisotropy in Homogeneous Shear Flows}}.  \jt{Phys. Rev. Lett.}
  \bvol{75},  \pg{3114--3117}.

\bibitem[Sagaut(2006)]{sagaut2006large}
{\sc \au{Sagaut, Pierre}} \yr{2006} {\em Large eddy simulation for
  incompressible flows: an introduction\/}.  \publ{Springer Science \& Business
  Media}.

\bibitem[Samiee {\em et~al.\/}(2020)Samiee, Akhavan-Safaei \&
  Zayernouri]{samiee2020fractional}
{\sc \au{Samiee, Mehdi}, \au{Akhavan-Safaei, Ali} \& \au{Zayernouri, Mohsen}}
  \yr{2020}  \at{A fractional subgrid-scale model for turbulent flows:
  Theoretical formulation and a priori study}.  \jt{Physics of Fluids}
  \bvol{32}~(5),  \pg{055102}.

\bibitem[Samiee {\em et~al.\/}(2021)Samiee, Akhavan-Safaei \&
  Zayernouri]{samiee2021tempered}
{\sc \au{Samiee, Mehdi}, \au{Akhavan-Safaei, Ali} \& \au{Zayernouri, Mohsen}}
  \yr{2021}  \at{{Tempered Fractional LES Modeling}}.  \jt{arXiv preprint
  arXiv:2103.01481} .

\bibitem[Schumacher {\em et~al.\/}(2005)Schumacher, Sreenivasan \&
  Yeung]{schumacher2005very}
{\sc \au{Schumacher, Joerg}, \au{Sreenivasan, Katepalli~R} \& \au{Yeung, PK}}
  \yr{2005}  \at{Very fine structures in scalar mixing}.  \jt{Journal of Fluid
  Mechanics}  \bvol{531},  \pg{113--122}.

\bibitem[Shraiman \& Siggia(1994)]{shariman_1994}
{\sc \au{Shraiman, Boris~I.} \& \au{Siggia, Eric~D.}} \yr{1994}  \at{Lagrangian
  path integrals and fluctuations in random flow}.  \jt{Phys. Rev. E}
  \bvol{49},  \pg{2912--2927}.

\bibitem[Shraiman \& Siggia(2000)]{shraiman2000scalar}
{\sc \au{Shraiman, Boris~I} \& \au{Siggia, Eric~D}} \yr{2000}  \at{Scalar
  turbulence}.  \jt{Nature}  \bvol{405}~(6787),  \pg{639--646}.

\bibitem[Song \& Karniadakis(2021)]{song2021variable}
{\sc \au{Song, Fangying} \& \au{Karniadakis, George~Em}} \yr{2021}
  \at{{Variable-Order Fractional Models for Wall-Bounded Turbulent Flows}}.
  \jt{Entropy}  \bvol{23}~(6),  \pg{782}.

\bibitem[Sreenivasan(1996)]{sreenivasan1996passive}
{\sc \au{Sreenivasan, Katepalli~R}} \yr{1996}  \at{{The passive scalar spectrum
  and the Obukhov--Corrsin constant}}.  \jt{Physics of Fluids}  \bvol{8}~(1),
  \pg{189--196}.

\bibitem[Sreenivasan(2019)]{sreenivasan2019turbulent}
{\sc \au{Sreenivasan, Katepalli~R}} \yr{2019}  \at{{Turbulent mixing: A
  perspective}}.  \jt{Proceedings of the National Academy of Sciences}
  \bvol{116}~(37),  \pg{18175--18183}.

\bibitem[Suzuki {\em et~al.\/}(2021{\natexlab{{\em a\/}}})Suzuki, Gulian,
  Zayernouri \& D'Elia]{suzuki2021fractional}
{\sc \au{Suzuki, Jorge}, \au{Gulian, Mamikon}, \au{Zayernouri, Mohsen} \&
  \au{D'Elia, Marta}} \yr{2021{\natexlab{{\em a\/}}}}  \at{{Fractional Modeling
  in Action: A Survey of Nonlocal Models for Subsurface Transport, Turbulent
  Flows, and Anomalous Materials}}.  \jt{arXiv preprint arXiv:2110.11531} .

\bibitem[Suzuki {\em et~al.\/}(2016)Suzuki, Zayernouri, Bittencourt \&
  Karniadakis]{suzuki2016fractional}
{\sc \au{Suzuki, JL}, \au{Zayernouri, M}, \au{Bittencourt, ML} \&
  \au{Karniadakis, GE}} \yr{2016}  \at{Fractional-order uniaxial
  visco-elasto-plastic models for structural analysis}.  \jt{Computer Methods
  in Applied Mechanics and Engineering}  \bvol{308},  \pg{443--467}.

\bibitem[Suzuki {\em et~al.\/}(2021{\natexlab{{\em b\/}}})Suzuki, Zhou,
  D’Elia \& Zayernouri]{suzuki2021thermodynamically}
{\sc \au{Suzuki, Jorge}, \au{Zhou, Yongtao}, \au{D’Elia, Marta} \&
  \au{Zayernouri, Mohsen}} \yr{2021{\natexlab{{\em b\/}}}}  \at{A
  thermodynamically consistent fractional visco-elasto-plastic model with
  memory-dependent damage for anomalous materials}.  \jt{Computer Methods in
  Applied Mechanics and Engineering}  \bvol{373},  \pg{113494}.

\bibitem[Suzuki {\em et~al.\/}(2021{\natexlab{{\em c\/}}})Suzuki, Kharazmi,
  Varghaei, Naghibolhosseini \& Zayernouri]{suzuki2021anomalous}
{\sc \au{Suzuki, Jorge~L}, \au{Kharazmi, Ehsan}, \au{Varghaei, Pegah},
  \au{Naghibolhosseini, Maryam} \& \au{Zayernouri, Mohsen}}
  \yr{2021{\natexlab{{\em c\/}}}}  \at{Anomalous nonlinear dynamics behavior of
  fractional viscoelastic beams}.  \jt{Journal of Computational and Nonlinear
  Dynamics}  \bvol{16}~(11),  \pg{111005}.

\bibitem[Suzuki {\em et~al.\/}(2021{\natexlab{{\em d\/}}})Suzuki, Tuttle,
  Roccabianca \& Zayernouri]{suzuki2021UBT}
{\sc \au{Suzuki, Jorge~L}, \au{Tuttle, Tyler~G}, \au{Roccabianca, Sara} \&
  \au{Zayernouri, Mohsen}} \yr{2021{\natexlab{{\em d\/}}}}  \at{{A Data-Driven
  Memory-Dependent Modeling Framework for Anomalous Rheology: Application to
  Urinary Bladder Tissue}}.  \jt{arXiv preprint arXiv:2110.00134} .

\bibitem[Waleffe(1992)]{waleffe1992nature}
{\sc \au{Waleffe, Fabian}} \yr{1992}  \at{The nature of triad interactions in
  homogeneous turbulence}.  \jt{Physics of Fluids A: Fluid Dynamics}
  \bvol{4}~(2),  \pg{350--363}.

\bibitem[Warhaft(2000)]{warhaft2000passive}
{\sc \au{Warhaft, Zellman}} \yr{2000}  \at{Passive scalars in turbulent flows}.
   \jt{Annual Review of Fluid Mechanics}  \bvol{32}~(1),  \pg{203--240}.

\bibitem[Watanabe \& Gotoh(2006)]{watanabe2006intermittency}
{\sc \au{Watanabe, Takeshi} \& \au{Gotoh, Toshiyuki}} \yr{2006}
  \at{Intermittency in passive scalar turbulence under the uniform mean scalar
  gradient}.  \jt{Physics of Fluids}  \bvol{18}~(5),  \pg{058105}.

\bibitem[Yaglom(1949)]{yaglom1949local}
{\sc \au{Yaglom, A.~M.}} \yr{1949} On the local structure of a temperature
  field in a turbulent flow.  \bt{In {\em Dokl. Akad. Nauk. SSSR\/}}, ,
  \vol{vol.~69}.

\bibitem[Yu {\em et~al.\/}(2016)Yu, Perdikaris \&
  Karniadakis]{yu2016fractional}
{\sc \au{Yu, Yue}, \au{Perdikaris, Paris} \& \au{Karniadakis, George~Em}}
  \yr{2016}  \at{Fractional modeling of viscoelasticity in 3d cerebral arteries
  and aneurysms}.  \jt{Journal of computational physics}  \bvol{323},
  \pg{219--242}.

\end{thebibliography}

\end{document}